\documentclass[twoside,preprintnumbers,amsmath,amssymb, nofootinbib]{revtex4}
\usepackage{epsfig, subfigure}
\usepackage{mathpazo}
\usepackage{graphicx}
\usepackage{dcolumn}
\usepackage{bm}
\usepackage{braket}
\usepackage[small,bf]{caption}

\usepackage{amsthm} 

\newcommand{\half}{\mbox{\small{$\frac{1}{2}$}}}

\newcommand{\MSbar}{\overline{\mbox{MS}}}

\newcommand{\occ}{\overline{c}}

\newcommand{\p}{\partial}

\newcommand{\wsigma}{\widetilde{\sigma}}
\newcommand{\e}{\ensuremath{\mathrm{e}}}
\newcommand{\en}{\ensuremath{\mathrm{en}}}
\newcommand{\tot}{\ensuremath{\mathrm{tot}}}
\newcommand{\GZ}{\ensuremath{\mathrm{GZ}}}
\renewcommand{\d}{\ensuremath{\mathrm{d}}}
\renewcommand{\dim}{\ensuremath{\mathrm{dim}}}
\newcommand{\lco}{\ensuremath{\overline{\varphi}\varphi-\overline{\omega}\omega}}

\begin{document}
\title{{\Large  The Landau gauge gluon and ghost propagator in the refined Gribov-Zwanziger framework in 3 dimensions} }

\author{D.~Dudal$^{a,b}$}
    \email{ddudal@mit.edu,david.dudal@ugent.be}

\author{J.A.~Gracey$^c$}
    \email{gracey@liv.ac.uk}

\author{S.P.~Sorella$^d$}
    \email{sorella@uerj.br}
    \altaffiliation{Work supported by FAPERJ, Funda{\c c}{\~a}o de Amparo {\`a} Pesquisa do Estado do Rio de Janeiro,
                                    under the program {\it Cientista do Nosso Estado}, E-26/100.615/2007.}
\author{N.~Vandersickel$^b$}
    \email{nele.vandersickel@ugent.be}

\author{H.~Verschelde$^b$}
 \email{henri.verschelde@ugent.be}
 \affiliation{\vskip 0.1cm
                            $^a$ Center for Theoretical Physics, Massachusetts Institute of Technology\\
                            77 Massachusetts Avenue, Cambridge, MA 02139, USA \\\\\vskip 0.1cm
                            $^b$ Ghent University, Department of Mathematical Physics and Astronomy \\
                            Krijgslaan 281-S9, B-9000 Gent, Belgium\\\\\vskip 0.1cm
                            $^c$ Theoretical Physics Division, Department of Mathematical Sciences, University of Liverpool\\ P.O. Box 147, Liverpool, L69 3BX, United Kingdom\\\\\vskip 0.1cm
                            $^d$ Departamento de F\'{\i }sica Te\'{o}rica, Instituto de F\'{\i }sica, UERJ - Universidade do Estado do Rio de Janeiro\\
                            Rua S\~{a}o Francisco Xavier 524, 20550-013 Maracan\~{a}, Rio de Janeiro, Brasil
                             }


\begin{abstract}\noindent
In previous works, we have constructed a refined version of the
Gribov-Zwanziger action in 4 dimensions, by taking into account a
novel dynamical effect. In this paper, we explore the 3-dimensional
case. Analogously as in 4 dimensions, we obtain a ghost propagator
behaving like $1/p^2$ in the infrared, while the gluon propagator
reaches a finite nonvanishing value at zero momentum.
Simultaneously, a clear violation of positivity by the gluon
propagator is also found. This behaviour of the propagators turns
out be in agreement with the recent numerical simulations.
\end{abstract}

\preprint{MIT-CTP 3969}\preprint{LTH-800} \maketitle
\setcounter{page}{1}

\section{Introduction}
Lately, the infrared behavior of the gluon and the ghost propagator
in $SU(N)$ Yang-Mills theories has been exhaustively investigated by
many research groups. The low energy behavior of these propagators
are of great interest as they might provide helpful information on
various aspects of color confinement, which is still far from being
understood. Previous results on the gluon and ghost propagators in
the Landau gauge have reported an enhanced behavior for the ghost
and a suppressed gluon propagator vanishing at zero momentum. This
behavior was supported by both numerical simulations
\cite{Cucchieri:2004mf,Sternbeck:2004xr} and analytical studies
\cite{Alkofer:2000wg,Lerche:2002ep,Pawlowski:2003hq,Alkofer:2003jj,Gribov:1977wm,Zwanziger:1989mf,Zwanziger:1992qr,Zwanziger:2001kw,Gracey:2005cx}.
\\\\ Nevertheless, recent lattice data on bigger
volumes point towards a ghost propagator which is no longer enhanced
and a gluon propagator which attains a finite value at zero momentum
\cite{Cucchieri:2007rg,Cucchieri:2007md,Bogolubsky:2007ud,Cucchieri:2008fc}.
Recently, several analytical approaches
\cite{Boucaud:2008ky,Aguilar:2008xm,Dudal:2007cw,Dudal:2008sp,Matevosyan:2008nf}
have been worked out, being in agreement with these new
data.  \\

Among these approaches, we have settled for a framework within the
Gribov-Zwanziger approach, successfully employed in the study of the
propagators in $4D$ \cite{Dudal:2007cw,Dudal:2008sp} in the Landau
gauge, which is defined as $\p_\mu A_\mu=0$. We recall here that
this condition does not uniquely fix the gauge freedom, as there can
be configurations $A_\mu'$, gauge equivalent to $A_\mu$, which also
fulfill $\p_\mu A_\mu'=0$ \cite{Gribov:1977wm}. The Gribov-Zwanziger
action allows for a (partial) resolution of this problem of gauge
(Gribov) copies in a local and renormalizable setting
\cite{Gribov:1977wm,Zwanziger:1989mf,Zwanziger:1992qr}. \\

Our refined framework was constructed as follows. We have added two
extra terms to the ordinary Gribov-Zwanziger action, without
destroying the renormalizability of the action. Let us already
recall here that these additional terms precisely correspond to
extra, yet unexplored, dynamical effects associated with the
Gribov-Zwanziger action \cite{Dudal:2007cw,Dudal:2008sp}. The first
extra term corresponded to the introduction of a novel mass
operator, while the second term represented an additional vacuum
 energy term which was required in order to remain
within the Gribov region $\Omega$. This region is defined as the set
of field configurations fulfilling the Landau gauge condition,
 $\partial_{\mu}A^{a}_{\mu}=0$, and for which the
Faddeev-Popov operator,
\begin{eqnarray}
\mathcal{M}^{ab} &=&  -\partial_{\mu}\left( \p_{\mu} \delta^{ab} + g
f^{acb} A^c_{\mu} \right) \;, \label{faddeev}
\end{eqnarray}
is strictly positive, namely
\begin{eqnarray}
\Omega &\equiv &\{ A^a_{\mu}, \; \partial_{\mu} A^a_{\mu}=0, \;
\mathcal{M}^{ab}  >0\} \;. \label{omega}
\end{eqnarray}
This region is bounded by the horizon, $\p \Omega$, where the first
vanishing eigenvalue of $\mathcal{M}^{ab}$ appears. Doing so, a
large number of gauge copies is already excluded, as their
appearance is precisely related to the existence of zero modes for
$\mathcal{M}^{ab}$ \cite{Gribov:1977wm}. So far, we have only worked
out the 4-dimensional case. The implementation of the
Gribov-Zwanziger approach in $3D$ has not yet been carried out. The
$3D$ case is also conceptually different from the $4D$ case due to
the superrenormalizability, and there is no running of the coupling
constant $g^2$.  Hence, it is useful to study the $3D$ case in
detail and make a comparison with
the $4D$ case as well as with the available $3D$ lattice data.\\

Also in lattice simulations, one has to deal with the existence of
gauge copies in the Landau gauge when studying the propagators. The
Landau gauge is numerically implemented by minimizing the functional
$\int \d^3x\;A^2$ over its gauge orbit. Notice that the Gribov
region $\Omega$ corresponds in fact to the set of all relative
minima of  $\int \d^3x\;A^2$. An ideal implementation of the Landau
gauge would correspond to finding the absolute minima of that
functional. Even numerically, this is an extremely hard task to
achieve, let stand alone to do it in the continuum formulation.
Nevertheless, the gauge field configurations employed in the
numerical evaluation correspond to relative minima of the functional
$\int \d^3x\;A^2$, so that they belong to the region $\Omega$. Let
us also mention that, at present,  it is unknown how to restrict in
the continuum the path integration to gauge fields corresponding to
absolute minima of $\int \d^3x\;A^2$. That set of absolute minima
forms a subset of the Gribov region $\Omega$, known as the
fundamental modular region $\Lambda$, $\Lambda\subset\Omega$. This
means that $\Omega$ itself is still plagued by additional gauge
copies \cite{Semenov,Dell'Antonio:1991xt,Dell'Antonio2}. Albeit it
has been argued in \cite{Zwanziger:2003cf} that these additional
copies should not contribute to the correlation functions, so that
there should be no difference between the regions $\Omega$ and
$\Lambda$, we point out that the restriction to $\Omega$ via the
local and renormalizable Gribov-Zwanziger approach is the best one
can do for now in the continuum. Summarizing, both in the lattice
and continuum study of the gauge theory, one is looking at the same
objects, e.g. gluon and ghost propagator (see Appendix C and
Section V), in such a way that the issue of gauge copies
has been at least partially handled. We refer to e.g. \cite{Cucchieri:2007md,Bogolubsky:2007ud,Cucchieri:2007rg,Cucchieri:2008fc,Cucchieri:2008yp} for lattice works.\\

For the benefit of the reader, we review the original construction
of the Gribov-Zwanziger action in section II. Section III presents a
detailed motivation of why we should include effects of a mass term
like $ S_{\overline{\varphi} \varphi} = M^2 \int \d^3
x\left(\overline{\varphi}\varphi-\overline{\omega}\omega\right)$ to
the Gribov-Zwanziger action $S_{\GZ}$. Also, the renormalizability
of $S_{GZ} + S_{\overline{\varphi} \varphi}$ is discussed. The
requirement of renormalizability/consistency with the symmetries of
the model puts rather severe restrictions on the possible additional
operators which can be introduced in the theory. We also provide
some details on the breaking of the BRST symmetry. In section IV we
discuss the need of the inclusion of an extra vacuum term $S_{\en} =
M^2 \int \d^3x \frac{2 (N^2 -1)}{ g^2 N} \varsigma  \lambda^2 $.
Again, we prove the renormalizability of the total action $S_{\tot}
= S_{\GZ} + S_{\overline{\varphi} \varphi} + S_{\en}$. In section V,
we investigate in detail the gluon and ghost propagator and already
show that for $M^2 \not = 0$, the ghost propagator is not enhanced
and the gluon propagator is finite at zero momentum, in accordance
with the lattice data of
\cite{Cucchieri:2007rg,Cucchieri:2007md,Bogolubsky:2007ud,Cucchieri:2008fc}.
In Section VI a variational technique is scrutinized in order to
determine the value of $M^2$. This mass parameter $M^2$ is thus not
added by hand to the original Gribov-Zwanziger action, but is
determined in a self-consistent way. Using this variational
technique, we shall also investigate the one loop ghost propagator
in the full momentum range. Finally, our conclusion is given in
section VII.

\section{The original Gribov-Zwanziger action}
We summarize the construction of the Gribov-Zwanziger action in $d$
dimensions, where it is understood that we actually work in
dimensional regularization, with $d=3-2\varepsilon$. We start from
the following action \cite{Zwanziger:1992qr}
\begin{equation}\label{horizon}
S_{\mathrm{start}}=S_{\mathrm{YM}}+ S_{\mathrm{Landau}} -
\gamma^{4}g^{2}\int \d^{d}xf^{abc}A_{\mu }^{b}\left(
\mathcal{M}^{-1}\right) ^{ad}f^{dec}A_{\mu }^{e} \;,
\end{equation}
where
\begin{equation}
S_{\mathrm{YM}} = \frac{1}{4}\int \d^d x F^a_{\mu\nu}F^a_{\mu\nu}
\end{equation}
is the classical Yang-Mills action and
\begin{equation}\label{landau}
S_{\mathrm{Landau}}= \int \d^d x (b^a\partial_{\mu}A^a_{\mu} +
\overline{c}^a \partial_{\mu} D^{ab}_{\mu} c^b)
\end{equation}
denotes the Landau gauge fixing and the ghost part. The part of
 expression \eqref{horizon} proportional to the
 Gribov mass parameter $\gamma$ is the nonlocal horizon
function, which implements the restriction to the first Gribov
region, with the proviso that this $\gamma$ is not free, but subject
to the horizon condition \cite{Zwanziger:1992qr}
\begin{equation}
\braket{h(x)}=d(N^2-1) \;, \label{hzoncond}
\end{equation}
where $h(x)$ is the so called horizon function
\begin{equation}
h(x) = g^{2} f^{abc}A_{\mu }^{b}\left( \mathcal{M}^{-1}\right)
^{ad}f^{dec}A_{\mu }^{e} \;. \label{hz}
\end{equation}
This nonlocal horizon function has also received attention
from the lattice community, see for example \cite{Furui:2005bu},
where a lattice formulation of the horizon function can be found.\\

In order to find a manageable local quantum field theory, one
 adds extra fields
$\left(\overline{\varphi}_{\mu}^{ac},\varphi_{\mu}^{ac},\overline{\omega}_{\mu
}^{ac},\omega_{\mu}^{ac}\right)$ to localize the nonlocal part of
the action \eqref{horizon}. Doing so, the Gribov-Zwanziger action
becomes \cite{Zwanziger:1992qr,Dudal:2005}
\begin{equation}
S_{\GZ}=S_{0}-\int \d^{d}x\left(\gamma
^{2}gf^{abc}A_{\mu}^{a}\varphi _{\mu }^{bc}+\gamma
^{2}gf^{abc}A_{\mu}^{a}\overline{\varphi }_{\mu }^{bc} + d\left(
N^{2}-1\right)\gamma^{4} \right)\;, \label{GribovZwanziger}
\end{equation}
with
\begin{eqnarray}
S_{0} &=&S_{\mathrm{YM}}+\int \d^{d}x\;\left( b^{a}\partial_\mu A_\mu^{a}+\overline{c}^{a}\partial _{\mu }\left( D_{\mu }c\right) ^{a}\right) \;  \nonumber \\
&+&\int \d^{d}x\Bigl( \overline{\varphi }_{i}^{a}\partial
_{\nu}\left( D_{\nu }\varphi _{i}\right)
^{a}-\overline{\omega}_{i}^{a}\partial _{\nu}\left( D_{\nu }\omega
_{i}\right) ^{a}  -g\left( \partial _{\nu }\overline{\omega
}_{i}^{a}\right) f^{abm}\left( D_{\nu }c\right)^{b}\varphi
_{i}^{m}\Bigr)  \;,
\end{eqnarray}
whereby $\left( \overline{\varphi }_{\mu
}^{ac},\varphi_{\mu}^{ac}\right) $ are a pair of complex conjugate
bosonic fields, whereas  $\left( \overline{\omega }_{\mu
}^{ac},\omega_{\mu}^{ac}\right)$ are anticommuting ghost field.
Based on a global $U(f)$ symmetry, $f=d\left( N^{2}-1\right)$, with
respect to the composite index $i=\left( \mu ,c\right)$ of the
additional fields $\left( \overline{ \varphi }_{\mu }^{ac},\varphi
_{\mu}^{ac},\overline{\omega }_{\mu}^{ac},\omega _{\mu
}^{ac}\right)$, we have introduced a more convenient notation
$\left( \overline{\varphi}_{\mu }^{ac},\varphi _{\mu
}^{ac},\overline{\omega}_{\mu }^{ac},\omega _{\mu }^{ac}\right)
=\left( \overline{\varphi }_{i}^{a},\varphi
_{i}^{a},\overline{\omega }_{i}^{a},\omega_{i}^{a}\right)$. Notice
that for $d=3$, $\dim(g^2)=1$ and $\dim(\gamma^2)=3/2$. Defining the
quantum effective action $\Gamma$ by means of
\begin{equation}\label{quantact}
    \e^{-\Gamma}=\int \d\Psi \e^{-S} \;,
\end{equation}
where $\Psi$ is a shorthand for all the fields, then it is easily
shown that the horizon condition \eqref{hzoncond} is in fact
equivalent with the requirement that
\begin{equation}\label{horizoncond2}
    \frac{\p \Gamma}{\p \gamma^2}=0
\end{equation}
is fulfilled for $\gamma^2\neq0$, or
\begin{equation}\label{horizoncond}
\Braket{
gf^{abc}A_\mu^a(\varphi_{\mu}^{bc}+\overline{\varphi}_{\mu}^{bc})}=-2d\left(N^2-1\right)\gamma^2
\;.
\end{equation}
This means that $\gamma$ will become fixed in terms of the natural
scale of the theory which, in $3D$, is provided
by the coupling itself. We thus expect to find $\gamma^2\propto g^3$.\\

As it has been shown in
\cite{Zwanziger:1992qr,Dudal:2005,Maggiore:1993wq}, the action
\eqref{GribovZwanziger} is renormalizable to all orders, and is thus
suitable for quantum computations.

\section{The Gribov-Zwanziger action complemented with a new operator}
\subsection{Proposal}
 As carried out in the 4-dimensional case
\cite{Dudal:2007cw,Dudal:2008sp}, we introduce the local composite
operator (LCO) $\overline{\varphi}\varphi$ into the action
\eqref{GribovZwanziger}. We shall couple this mass term to the
action using a source $J$, and for renormalization purposes, we have
to add this term in a BRST invariant way. Therefore, we shall see
that the $\omega$-sector must gain the same mass. Hence, we
 consider the following action:
\begin{eqnarray}\label{extendedaction}
    \label{nact}S' &=& S_{GZ} + S_{\overline{\varphi} \varphi}, \nonumber\\
    S_{\overline{\varphi} \varphi} &=& \int \d^d x \left[-J\left( \overline{\varphi}^a_i \varphi^a_{i} - \overline{\omega}^a_i \omega^a_i \right) \right]   \;, \label{s3eq1}
\end{eqnarray}
with $S_{GZ}$ the original Gribov-Zwanziger action
\eqref{GribovZwanziger}. The new source $J$ has the dimension of a
mass squared. Therefore, we shall often use the notation $M^2 = J$.

\subsection{Motivation}
Let us briefly repeat the motivation for adding an extra term to the
Gribov-Zwanziger action. As pointed out in \cite{Dudal:2007cw}, the
fields $\left( \overline{\varphi }_{\mu }^{ac},\varphi _{\mu
}^{ac},\overline{\omega}_{\mu}^{ac},\omega _{\mu }^{ac}\right)$
introduced to localize the horizon function appearing in
\eqref{horizon} are interacting fields corresponding in fact to the
nonlocal dynamics associated with the horizon function. Therefore,
these fields will develop their own quantum dynamics. We draw
attention to the fact that the $A$- and $\varphi$- fields are
intimately entangled, since there is a quadratic $A\varphi$-mixing
term present in the tree level action \eqref{GribovZwanziger},
namely $g\gamma^2 f^{abc}A^{a}_{\mu}(\varphi _{\mu }^{bc}+
\overline{\varphi }_{\mu }^{bc})$. Hence, we expect that any effect
in the $\varphi$-sector will immediately reflect in the gluon
sector, altering, in particular, the behavior of the gluon
propagator at zero momentum. This is nothing else
than saying that the horizon strongly influences the gluon dynamics.\\

On top of the previous argument, we can give a second argument why
one should add the local composite operator (LCO)
$(\overline{\varphi}\varphi-\overline{\omega}\omega)$ to the action.
The horizon condition corresponds to
\begin{eqnarray}
\Braket{gf^{abc}A_\mu^a(\varphi_{\mu}^{bc}
+\overline{\varphi}_{\mu}^{bc})}=-2d\left(N^2-1\right)\gamma^2 \;,
\end{eqnarray}
i.e. to a condensate proportional to $\gamma^2$. One might then also
expect that
$\braket{\overline{\varphi}\varphi-\overline{\omega}\omega}$ will be
nonvanishing too when $\gamma^2\neq0$. Hence, one is almost obliged
to incorporate the effects related to the operator
$\overline{\varphi}\varphi-\overline{\omega}\omega$. To make this
argument more explicit, let us compute the \emph{perturbative} value
of the condensate
$\langle\overline{\varphi}\varphi-\overline{\omega}\omega\rangle$.
We start from
\begin{equation}\label{pot4}
    \braket{\overline{\varphi}\varphi-\overline{\omega}\omega}_{\mathrm{pert}}=-\left.\frac{\p W(J)}{\p
    J}\right\vert_{J=0} \;,
\end{equation}
with $W(J)$ the generating functional defined in our case as
\begin{eqnarray}\label{genfunc}
\e^{-W(J)} &=& \int [\d \Psi] \e^{-S'}
\end{eqnarray}
and with $S'$ the extended Gribov-Zwanziger action given in
\eqref{extendedaction}. At lowest order, one finds
\begin{equation}\label{pot1}
    W(J)=-d(N^2-1)\gamma^4+\frac{N^2-1}{2}(d-1)\int \frac{\d^d q}{(2\pi)^d}\ln\left(q^4+q^2\frac{2g^2N\gamma^4}{q^2+J}\right) \;.
\end{equation}
Making use of
\begin{eqnarray}
   \int \frac{\d^d
    q}{(2\pi)^d}\ln\left(q^4+q^2\frac{2g^2N\gamma^4}{q^2+J}\right)&=& \int \frac{\d^d
    q}{(2\pi)^d}\ln(q^4+q^2J+2g^2N\gamma^4)-\int \frac{\d^d
    q}{(2\pi)^d}\ln(q^2+J)\nonumber\\
    &=&\int \frac{\d^d
    q}{(2\pi)^d}\ln(q^2+\omega_+^2)+\int \frac{\d^d
    q}{(2\pi)^d}\ln(q^2+\omega_-^2)-\int \frac{\d^d
    q}{(2\pi)^d}\ln(q^2+J)
\end{eqnarray}
with
\begin{equation}
\omega_{\pm}^2=\frac{J\pm\sqrt{J^2-4\lambda^4}}{2} \;,
\end{equation}
where $\lambda^4 = 2 g^2 N \gamma^4$, and after employing a standard
integral in dimensional regularization,
\begin{equation}\label{pot3}
    \int \frac{\d^d     \ell}{(2\pi)^d}\ln(\ell^2+m^2)=-\frac{(m^2)^{d/2}}{(4\pi)^{d/2}}\Gamma(-d/2) \;,
\end{equation}
 we find the following ultraviolet \emph{finite} result
\begin{equation}\label{pot1}
    W(J)=-3(N^2-1)\frac{\lambda^4}{2g^2N}+\frac{N^2-1}{6\pi}\left(-\omega_+^3-\omega_-^3+J^{3/2}\right) \;.
\end{equation}
Using this explicit expression, we can easily obtain the
perturbative value of the condensate \eqref{pot4}, reading
\begin{equation}\label{pot5}
\braket{\overline{\varphi}\varphi-\overline{\omega}\omega}_{\mathrm{pert}}=\sqrt{2}\frac{N^2-1}{8\pi}\lambda\approx
0.056(N^2-1)\lambda\;,
\end{equation}
where $\lambda$ is the nonzero solution of
$\frac{\p\Gamma(\lambda)}{\p \lambda}=0$. Since at one loop
\begin{eqnarray}\label{gribovpuur}
    \Gamma(\lambda)&=&-d(N^2-1)\frac{\lambda^4}{2Ng^2}+\frac{N^2-1}{2}(d-1)\int \frac{\d^d
    q}{(2\pi)^d}\ln\left(q^4+\lambda^4 \right)\nonumber\\
&=&-3(N^2-1)\frac{\lambda^4}{2g^2N}+\frac{\sqrt{2}}{6\pi}(N^2-1)\lambda^3
\;,
\end{eqnarray}
we find that
\begin{equation}\label{gribovpuuropl}
    \lambda=\frac{\sqrt{2}}{12\pi}g^2N \;,
\end{equation}
with
\begin{equation}\label{gribovpuurvac}
    E_{\mathrm{vac}}=g^6\frac{N^3(N^2-1)}{10368\pi^4}>0 \;.
\end{equation}
We notice that the one loop vacuum energy corresponding to the
Gribov-Zwanziger action is \emph{positive}. The same feature was
also observed in the one
loop $4D$ case \cite{Dudal:2008sp}.\\

Besides this perturbative value of the condensate \eqref{pot5}, it
could also be possible that another, non-perturbative, value
emerges. However, to calculate this non-perturbative value, one
should calculate the Legendre transformation of the generating
functional $W(J)$ to obtain the effective action $\Gamma(\sigma)$,
where the absolute minimum configuration
$\sigma_*\sim\braket{\overline{\varphi}\varphi-\overline{\omega}\omega}$
would correspond to the true, energetically favoured, vacuum.
Here, we limit ourselves to observe that, similar to what was
encountered in $4D$, \cite{Dudal:2008sp},
 no such value has been found at one loop,
meaning that higher order contributions have to be taken into
account. For the interested reader, we have written down the
corresponding details in Appendix \ref{appendixA}.\\

\subsection{Renormalizability}
To prove the renormalizability, let us start from the following
action,
\begin{eqnarray}\label{nact}
    S'' &=& S' + S_{\mathrm{LCO}}  \;,  \nonumber\\
    S_{\mathrm{LCO}} &=& \int \d^d x \rho g^2 J\;,
\end{eqnarray}
with $S'$ the action proposed in \eqref{extendedaction}, and with
the LCO parameter $\rho$ a new dimensionless quantity. This term in
$\rho$ is, in principle, needed to take into account potential
divergences proportional to $g^2J$, which are allowed by power
counting and by the symmetries of the action\footnote{We refer to
\cite{Knecht:2001cc} for more details concerning the LCO formalism
and the precise role of the LCO parameter.}. This term would  ensure
multiplicative renormalizability of the functional $W(J)$. However,
although this term cannot be ruled out at the level of the algebraic
analysis, an analogous proof as in $4D$ can be written down allowing
us to consistently set $\rho=0$
\cite{Dudal:2008sp}.\\

To study the renormalizability of the Gribov-Zwanziger action, it is
highly useful to embed it in an extended action, which reduces to
the original model in a specific limit \cite{Zwanziger:1992qr}.
Doing so, we may have a larger number of Ward identities at our
disposal, which are powerful tools to construct the most general
possible counterterm. Therefore, proceeding as in the $4D$ case
\cite{Dudal:2008sp}, we shall start with the following action
\begin{eqnarray}\label{cact}
\Sigma
&=&S_{0}+S_{\mathrm{s}}+S_{\mathrm{ext}}+S_{\overline{\varphi}\varphi}
+ S_{\mathrm{LCO}}\;,
\end{eqnarray}
with $S_{0}$ given in \eqref{GribovZwanziger},
$S_{\overline{\varphi}\varphi}$ in \eqref{extendedaction}, and
\begin{eqnarray}
S_{\mathrm{s}} &=& s\int \d^{d}x\left( -U_{\mu }^{ai}\left( D_{\mu}\varphi _{i}\right) ^{a}-V_{\mu }^{ai}\left( D_{\mu }\overline{\omega }_{i}\right) ^{a}-U_{\mu}^{ai}V_{\mu }^{ai} \right)\;,\nonumber\\ &=& \int \d^{d}x\left( -M_{\mu }^{ai}\left( D_{\mu }\varphi _{i}\right) ^{a}-gU_{\mu }^{ai}f^{abc}\left( D_{\mu }c\right) ^{b}\varphi _{i}^{c}+U_{\mu }^{ai}\left( D_{\mu }\omega _{i}\right) ^{b}\right. \nonumber \\
&-&N_{\mu }^{ai}\left( D_{\mu }\overline{\omega }_{i}\right)
^{a}-V_{\mu }^{ai}\left( D_{\mu }\overline{\varphi }_{i}\right)
^{a}+gV_{\mu }^{ai}f^{abc}\left( D_{\mu }c\right)
^{b}\overline{\omega }_{i}^{c} -\left.M_{\mu }^{ai}V_{\mu
}^{ai}+U_{\mu }^{ai}N_{\mu }^{ai}\right) \;,
\nonumber\\S_{\mathrm{ext}}&=&\int \d^{d}x\left( -K_{\mu }^{a}\left(
D_{\mu }c\right)^{a}+\frac{1}{2}gL^{a}f^{abc}c^{b}c^{c}\right) \;.
\end{eqnarray}
We introduced new sources $M_{\mu}^{ai}$, $V_{\mu }^{ai}$
,$U_{\mu}^{ai}$, $N_{\mu}^{ai}$, $K_{\mu}^{a}$ and $L^{a}$, which
are necessary to analyze the renormalization of the corresponding
composite field operators in a BRST invariant fashion. The BRST
operator $s$ is defined through
\begin{align}\label{brst}
sA_{\mu }^{a} &=-\left( D_{\mu }c\right) ^{a}\;,  sc^{a}
=\frac{1}{2}gf^{abc}c^{b}c^{c}\;,  s\overline{c}^{a} =b^{a}\;,
sb^{a}=0\;,  s\varphi _{i}^{a} =\omega _{i}^{a}\;, s\omega
_{i}^{a}=0\;,  s\overline{\omega }_{i}^{a} =\overline{\varphi
}_{i}^{a}
\;, s \overline{\varphi }_{i}^{a} =0\;,\nonumber\\
sU_{\mu }^{ai} &= M_{\mu }^{ai}\;, sM_{\mu }^{ai}=0\;, sV_{\mu
}^{ai} = N_{\mu }^{ai}\;,  sN_{\mu }^{ai} =0\;, s K_{\mu}^a = 0,  s
L^{a} = 0\;, sJ=0\;,
\end{align}
 so that $s$ is nilpotent, $s^2 = 0$. One can easily
see that the action $\Sigma$ is indeed BRST invariant, $s\Sigma=0$ .
We underline that the mass operator itself,
$\overline{\varphi}\varphi-\overline{\omega}\omega$, is also BRST
invariant. In TABLE I, we have summarized for all the fields and
sources their mass dimension, ghost number and
$\mathcal{Q}_f$-charge, which is defined by means of the diagonal
generator $U_{ii}$ of the global $U(f)$ symmetry. We have chosen
these mass dimensions such that in any case, the action of the BRST
transformation $s$ raises the dimension
 by  $1/2$.\\
\begin{table}[t]
  \centering
        \begin{tabular}{|c|c|c|c|c|c|c|c|c|c|c|c|c|c|c|c|}
        \hline
        & $A_{\mu }^{a}$ & $c^{a}$ & $\overline{c}^{a}$ & $b^{a}$ & $\varphi_{i}^{a} $ & $\overline{\varphi }_{i}^{a}$ &                $\omega _{i}^{a}$ & $\overline{\omega }_{i}^{a}$&$U_{\mu}^{ai}$&$M_{\mu }^{ai}$&$N_{\mu }^{ai}$&$V_{\mu }^{ai}$&$K_{\mu }^{a}$&$L^{a}$&$J$ \\
        \hline
        \hline
        \textrm{dimension} & $1/2$ & $0$ &$1$ & $3/2$ & $1/2$ & $1/2$ & $1$ & $0$& $1$ & $3/2$ & $2$ &$3/2$  & $2$ & $5/2$ & $2$ \\
        \hline
        $\mathrm{ghost number}$ & $0$ & $1$ & $-1$ & $0$ & $0$ & $0$ & $1$ & $-1$& $-1$& $0$ & $1$ & $0$ & $-1$ & $-2$ &$0$ \\
        \hline
        $Q_{f}\textrm{-charge}$ & $0$ & $0$ & $0$ & $0$ & $1$ & $-1$& $1$ & $-1$& $-1$ & $-1$ & $1$ & $1$ & $0$ & $0$ &$0$\\
        \hline
        \end{tabular}
        \caption{Quantum numbers of the fields.}\label{tabel1}
        \end{table}

At the end, we can give the sources the following physical values
\begin{eqnarray}
&\left. M_{\mu \nu }^{ab}\right|_{\mathrm{phys}}= \left.V_{\mu \nu }^{ab}\right|_{\mathrm{phys}}=\gamma ^{2}\delta ^{ab}\delta _{\mu \nu}\;,& \nonumber\\
&\left. U_{\mu }^{ai}\right|_{\mathrm{phys}} = \left. N_{\mu
}^{ai}\right|_{\mathrm{phys}} =\left. K_{\mu
}^{a}\right|_{\mathrm{phys}} =\left. L^{a}\right|_{\mathrm{phys}}  =
0\;,&\label{physlim}
\end{eqnarray}
in order to recover the physically relevant action $S''$, given in
\eqref{nact}. If the action $\Sigma$ is renormalizable for any value
of the sources, it will of course be for the specific values
\eqref{physlim}. This is an example of the fact that it can be
highly useful to use a slightly more general version than the
original action, whereby a more powerful set of Ward identities can
be invoked to prove e.g. the renormalizability of $S'$. We have
collected the details of the algebraic renormalization analysis in
Appendix B. The main conclusion drawn is that the action $\Sigma$
\eqref{cact} is multiplicatively
renormalizable to all orders of perturbation theory.\\

It is important to notice here that it is the Ward identities
defining the extended Gribov-Zwanziger action \eqref{cact} which
dictate which terms can be added to the action, without jeopardizing
the symmetry content in general and the renormalizability in
particular. From this perspective, the mass term
\eqref{extendedaction} displays remarkable features. In fact, its
introduction turns out to be compatible with both Ward identities
and renormalizability, as originally proposed in
\cite{Dudal:2007cw}.

\subsection{The breaking of the BRST symmetry}
We would like to draw attention to the fact that the actions $S$ and
$S'$ are BRST invariant if $\gamma^2=0$, since then the extra fields
can be integrated out, and we are left with the original Yang-Mills
action in the Landau gauge. Evidently, we then have
$\braket{\overline{\varphi}\varphi-\overline{\omega}\omega}=-\braket{s(\overline{\omega}\varphi)}=0$.
It is only when $\gamma^2\neq0$ that we can have such a nonvanishing
condensate, since in that case the BRST symmetry generated by $s$ is
no longer preserved, since $sS'\neq0$. More precisely, the ``horizon
terms'' $\propto \gamma^2$ in equation \eqref{GribovZwanziger} are
not BRST invariant. It is an important observation that the
Gribov-Zwanziger action is \emph{not} BRST invariant: the BRST
symmetry is explicitly broken by soft terms $\sim \gamma^2$.
Nevertheless, it is worth emphasizing that this breaking can be kept
under control at the quantum level \cite{Dudal:2008sp}. In
particular, the introduction of the local sources $U$, $V$, $M$ and
$N$, see also \cite{Capri:2007ix}, nicely allows one to embed the
action into the larger BRST invariant action, thereby allowing one
to prove the renormalizability. When the sources attain their
physical values, eqs.(\ref{physlim}), the exact Slavnov-Taylor
identities of the larger action induce the corresponding softly
broken identities for the physical action $S''$. The operators
coupled to these sources are exactly those relevant for the
discussion of the broken Slavnov-Taylor identity; see a similar
discussion in \cite{Capri:2007ix}.\\

The introduction of the horizon function thus \emph{explicitly}
breaks the BRST invariance of the Landau gauge fixed Yang-Mills
action. In particular, in \cite{Dudal:2008sp}, it has been shown
that the origin of this breaking is deeply related to the
restriction to the Gribov region $\Omega$. More precisely, it turns
out that infinitesimal gauge transformations of field configurations
belonging to the Gribov region $\Omega$ give rise to configurations
lying outside of $\Omega$. The appearance of the BRST breaking looks
thus rather natural, when we keep in mind that the BRST
transformation of the gauge field is inherited from its
infinitesimal gauge transformation. Let us elaborate a bit on this
here. The starting point is that we must restrict the domain of path
integration to the region $\Omega$. Let us consider the full set of
fields present, $A_\mu$, $c$, $\occ$, $b$, and potential other
fields. These fields can be seen as the coordinates of a space $\cal
F$. We can define a manifold over $\cal F$, by means of the action
functional,
\begin{eqnarray} S: A_\mu, c, \occ, b,\ldots \in {\cal F}\to
S(A_\mu,c,\occ,b,\ldots)\in \mathbb{R}\,.
\end{eqnarray}
To be more precise, we must restrict ourselves to $\Omega$, so we
must consider the functional
\begin{eqnarray} S_{\mathrm{restricted}}: A_\mu, c, \occ,
b,\ldots \in {\cal F}\vert A_\mu\in\Omega\to
S(A_\mu,c,\occ,b,\ldots)\in \mathbb{R}\;.
\end{eqnarray}
In a pictorial way, we can imagine this as some kind of ``cylinder''
in $\cal F$ with $\Omega$ as ground surface, and the other
unrestricted fields ($c,\occ,\ldots$) describing the height. Notice
that we do not say what the action $S$ precisely is, we only assume
that it is BRST invariant. Now, we consider a particular point
$(A_\mu^*,c^*,\occ^*,b^*,\ldots)$ very close to the boundary of the
cylinder, thus $A_\mu^*$ is located very close to the inner boundary
of $\Omega$, and $\p_\mu A_\mu^*=0$. We can decompose $A_\mu^*$ as
\begin{eqnarray}\label{brstbreking4}
  A_\mu^* &=& a_\mu + C_\mu\;,
\end{eqnarray}
with $C_\mu \in \p\Omega$, thus $C_\mu$ lies on the Gribov horizon.
The shift $a_\mu$ is a small (infinitesimal) perturbation.
Obviously, $\p_\mu C_\mu = \p_\mu a_\mu=0$. We then find
\begin{eqnarray}\label{brstbreking5}
  \widetilde{A}_\mu &=& \underbrace{C_\mu+a_\mu}_{A_\mu^*}+D_\mu(C)\omega +\ldots
\end{eqnarray}
for the gauge transformed field. Since $C_\mu\in\p\Omega$ and
setting $\omega$ equal to the zero mode corresponding to $C_\mu$,
yields
\begin{eqnarray}\label{brstbreking6}
  \p_\mu\widetilde{A}_\mu &=& \p_\mu D_\mu(C)\omega~=~0\;.
\end{eqnarray}
In addition, $\widetilde{A}_\mu$ also lies very close to the
boundary $\p\Omega$, however on the other side, as shown by Gribov
in \cite{Gribov:1977wm}. Let us now consider the infinitesimal BRST
shift of the coordinate set $(A_\mu^*,c^*,\ldots)$, which is given
by
\begin{eqnarray}\label{brstbreking8}
A_\mu^*\to A_\mu^*+\theta D_\mu c^*
\end{eqnarray}
for the gauge field, with $\theta$ a Grassmann number. We did not
specify yet the choice of our particular ghost coordinate. We take
$c^*= \frac{\theta'}{\theta\theta'}\omega$, with $\theta'$ another
Grassmann number\footnote{Notice that $\theta\theta'$ is a normal
number, thus we can divide by it.}, and $\omega$ the zero mode.
Since the BRST was supposed to be a symmetry of the cylinder, the
transformed coordinate set of $(A_\mu^*,c^*,\ldots)$ should still be
located within the cylinder. However, by construction of $A_\mu^*$
and $c^*$, we do end up outside of the cylinder. This contradiction
means that maintaining the BRST symmetry is not possible when
restricting  to $\Omega$.\\

One can also imagine a configuration $A_\mu^{**}\in\Omega$
not located close the boundary $\p\Omega$, thus $\p_\mu
A_\mu^{**}=0$ and $-\partial_\mu D_\mu(A^{**})>0$. If we are to
assume that the BRST transformation of $A_\mu^{**}$,
\begin{eqnarray}
A_\mu^{**}+\theta D_\mu c^{**}\,,
\end{eqnarray}
with $c^{**}$ arbitrary, would remain within $\Omega$, we are forced
to conclude that $\p_\mu D_\mu[A^{**}] \left(\theta
c^{**}\right)=0$, thus $\theta c^{**}$ would then be a zero mode,
again in contradiction with the hypothesis that $A_\mu^{**}$ is not
located on (or close to) the boundary $\p\Omega$, meaning that there
are no such zero modes.\\

An interesting property worth being mentioned is that, despite the
loss of the BRST symmetry, one can still use the associated broken
Slavnov-Taylor identity to derive relations amongst several Green
functions. This has exhaustively been studied in
\cite{Dudal:2008sp}, and can be easily transported to the $3D$ case.
We therefore refer the reader to \cite{Dudal:2008sp} for any detail
concerning the BRST breaking and its consequences. In particular, we
have also algebraically motivated that the (controlled) BRST
breaking allows the Gribov parameter $\gamma$
to become a physical parameter. If the BRST symmetry would be preserved, $\gamma$ would merely play a role akin to that of an unphysical gauge parameter.\\

We emphasize here that the soft breaking of the BRST is introduced
in such a way that one keeps the nilpotency of the BRST operator.
This is an important point, as it enables one to make use of the
notion of the BRST cohomology. Indeed, as discussed earlier in this
section, the proof of the renormalizability of the action is exactly
possible by making use of the BRST cohomology in the extended model
\eqref{cact}, which enjoys the BRST symmetry. In fact, one can
define the physical operators in the Gribov-Zwanziger model as those
obtained from the physical operators $\mathcal{O}_{\mathrm{phys}}$
in the extended model \eqref{cact}, upon taking the physical limit
\eqref{physlim} of these operators $\mathcal{O}_{\mathrm{phys}}$,
which are nothing else than the cohomology classes of the nilpotent
BRST operator. The latter are precisely given by the gauge invariant
singlet operators constructed with the field strength and its
covariant derivative. In other words, due to the form of the BRST
operator, the physical operator content of the theory is left
unchanged, being identified with the colorless gauge invariant
operators. As a consequence, both gluons and ghosts are excluded
from the physical spectrum. Moreover, due to the presence of the
Gribov parameter $\gamma$ and of the mass $M$, the behavior of the
correlation function of gauge invariant operators will also get
modified in the infrared region, as it can be inferred from the
expression of the resulting gluon propagator. For instance, although
the gluon propagator turns out to exhibit positivity violations (see
section V, eq.(\ref{gluonprop2})), one might expect that the
correlation functions of gauge invariant operators, like e.g.
$\langle F^2(x) F^2(y) \rangle$, could display a real pole in
momentum space, which would be related to the mass of a glueball
bound state. This topic is currently under study. At asymptotically
large momenta, one can neglect the soft BRST breaking term, in which
case we are reduced to the normal Yang-Mills physical modes. The
degrees of freedom corresponding to the fields $\varphi$,
$\overline{\varphi}$, $\omega$ and $\overline{\omega}$ will decouple
from the physical spectrum, as these fields form BRST doublets, see
\eqref{brst}, and it is well known that these become trivial in the
BRST cohomology \cite{Piguet:1995er}.

\subsection{A few more words about the intricacies of $3D$ gauge theories: infrared problems and ultraviolet finiteness}
In the previous subsection, we mentioned the case $\gamma^2=0$.
Strictly speaking, the $3D$ theory will not be well defined in that
case. In the absence of an infrared regulator, the perturbation
theory of a super-renormalizable $3D$ gauge theory is ill-defined
due to severe infrared instabilities \cite{Jackiw:1980kv}.
This can be intuitively understood as the coupling constant
$g^2$ carries the dimension of a mass. In the absence of an infrared
regulator, the effective expansion parameter will look like $g^2/p$
with $p$ a certain (combination of) external momentum/-a. For $p\gg
g^2$, very good ultraviolet behaviour is apparent, but for $p\ll
g^2$, infrared problems emerge. The presence of (a) dynamical mass
scale(s) $m\propto g^2$ could ensure a sensible perturbation series,
even for small $p$, as a natural expansion parameter is then
provided by $g^2/m$. From this perspective, a nonvanishing Gribov
mass $\gamma^2$ could also serve as infrared cut-off. This feature
is also explicitly seen from eq.\eqref{gribovpuuropl}, from which an
effective dimensionless expansion parameter can be derived as
\begin{eqnarray}\label{exppar}
  \frac{g^2N}{(4\pi)^{3/2}\lambda} &=&\frac{3}{2\sqrt{2\pi}}\approx 0.6\;,
\end{eqnarray}
a quantity which is at least smaller than $1$. The inverse factor
$(4\pi)^{3/2}$ is the generic loop integration factor generated in $3D$.\\

Although this is not the main concern of this paper, it might be
interesting to perform higher order computations to effectively
find out whether all infrared divergences are absent when $\gamma^2\neq0$.\\

In fact, it is possible to couple the regulating mass term
$\frac{1}{2} m^2 A^2$ to the Gribov-Zwanziger action. We did not
consider this in the current paper, but in $4D$ this has been
discussed in full detail in \cite{Dudal:2008sp}. All Ward identities
and relations between renormalization constants are maintained.
Moreover, the form of the propagators is only quantitatively
influenced by this additional mass $m^2$, whereby the main
consequences of the mass related to
$\overline{\varphi}\varphi-\overline{\omega}\omega$ are preserved
(see later) \cite{Dudal:2008sp}. In the presence of $\frac{1}{2} m^2
A^2$, all $Z$-factors are 1, or said otherwise, there are no
ultraviolet divergences when computing Green functions
\cite{Dudal:2004ch,Dudal:2006ip}. Having a look at the relations
\eqref{renrel2} and \eqref{renrel}, this also means that any other
$Z$-factor is 1, and hence the Gribov-Zwanziger theory is completely
ultraviolet finite, including the vacuum functional, since there is
no independent renormalization for it: the potential divergences
related to $\gamma^4$ are killed by the already available
$Z$-factors, which are themselves trivial, and we already know that
there are no divergences related to $g^2J$. It is understood that,
if needed, the mass related to $A^2$ is brought back to zero, as the
other mass parameters are expected to cure the theory in the
infrared. This should be checked case by case. For the purposes of
this paper, we shall later see that everything works out fine
without a mass coupled to $A^2$.

\section{The Gribov-Zwanziger action supplemented by an extra vacuum term}
\subsection{Proposal}
We propose to add an extra vacuum term to the action
\eqref{extendedaction}, i.e.
\begin{eqnarray}\label{nieuweterm}
S_{\en} &=& 2 \frac{d (N^2 -1)}{\sqrt{2 g^2 N}}  \int \d^d x\
\varsigma \ \gamma^2 J \;.
\end{eqnarray}
where the prefactor $2 \frac{d (N^2 -1)}{\sqrt{2 g^2 N}}$ is chosen
for later convenience.

\subsection{Motivation}
\begin{figure}[t]
  \centering
      \includegraphics[width=14cm]{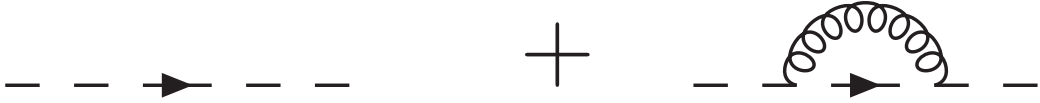}
  \caption{The one loop corrected ghost propagator.}  \label{ghostfigure}
\end{figure}
We shall now explain the need for the inclusion of this vacuum term.
So far, we have altered the original Gribov-Zwanziger action by
adding a mass operator to the action. However, we have to be careful
that with this addition we are not leaving the Gribov region when
performing calculations. Initially, staying inside the Gribov
horizon was assured by the horizon condition \eqref{horizoncond2}.
This condition is equivalent to demanding that the Faddeev-Popov
operator defined in \eqref{faddeev} is positive,
i.e.~$-\partial_{\mu} D^{ab}_{\mu} > 0$. In turn, looking at the
Landau gauge fixing \eqref{landau}, this is equivalent to demanding
the positivity of the inverse ghost propagator, which can be deduced
from FIG.~\ref{ghostfigure}, giving,
\begin{eqnarray} \label{ghostpropagator1}
\mathcal{G}^{ab}(k^2) &=&  \delta^{ab} \mathcal{G}(k^2) ~=~
\delta^{ab}\left( \frac{1}{k^2} + \frac{1}{k^2} \left[g^2
\frac{N}{N^2 - 1} \int \frac{\d^4 q}{(2\pi)^4} \frac{(k-q)_{\mu}
k_{\nu}}{(k-q)^2}
 \Braket{A^a_{\mu}A^a_{\nu}}\right] \frac{1}{k^2} \right) + \mathcal{O}(g^4) \nonumber\\
&=& \delta^{ab} \frac{1}{k^2} (1+ \sigma(k^2)) + \mathcal{O}(g^4)\;,
\end{eqnarray}
with
\begin{eqnarray}\label{ghi}
\sigma(k^2) &=& \frac{N}{N^2 - 1} \frac{g^2}{k^2}\int \frac{\d^4
q}{(2\pi)^4} \frac{(k-q)_{\mu} k_{\nu}}{(k-q)^2}
\Braket{A^a_{\mu}A^a_{\nu}} \;.
\end{eqnarray}
and with $\Braket{A^a_{\mu}A^a_{\nu}}$ the gluon propagator (see
later). We can now rewrite the ghost propagator by performing a resummation as
\begin{eqnarray}
\mathcal{G}(k^2) &=& \frac{1}{k^2} \frac{1}{1 - \sigma(k^2)} + O(g^4)
\end{eqnarray}
as we are only working up to order $g^4$. This corresponds
to the usual resummation of a set of connected diagrams into the
inverse of the one loop 1PI ghost self energy. We can therefore
write the inverse ghost propagator as,
\begin{eqnarray}
\mathcal{G}^{-1}(k^2) &=& k^2 (1 - \sigma(k^2)) +  O(g^4)\;,
\end{eqnarray}
Demanding the positivity of the previous expression is translated in
the so called Gribov no-pole condition \cite{Gribov:1977wm},
\begin{eqnarray}\label{nopole}
\sigma(k^2) \leq 1 \;,
\end{eqnarray}
which is the key-point of this investigation. Without the inclusion
of this extra vacuum term, we shall demonstrate that is impossible
to satisfy the no-pole condition. We recall that this no-pole
condition was the basis of the original Gribov paper \cite{Gribov:1977wm}.\\

We have introduced a new parameter $\varsigma$ which is still free
and needs to be determined. We notice that this vacuum term is,
 in a sense, comparable to the vacuum term already
present in the action \eqref{GribovZwanziger}, i.e.~$- \int \d^{d}x
d\left( N^{2}-1\right)\gamma^{4}$. In fact, in the original
Gribov-Zwanziger formulation, this term was also necessary to stay
within the horizon. Analogously as $\gamma$ is fixed by a gap
equation, we shall introduce a second gap equation to determine
$\varsigma$. Proceeding as in the $4D$ case, we shall impose that
\begin{eqnarray} \label{bound}
\left.\frac{\partial \sigma(0)}{\partial M^2} \right|_{M^2=0}&=&
0\;,
\end{eqnarray}
which assures a smooth limit to the original Gribov-Zwanziger case
when $M^2 \to 0$.

\subsection{Renormalizability}
The renormalizability of the following action,
\begin{eqnarray}\label{totaleactie}
S_{\tot} &=& S' + S_{\en} \nonumber\\
&=& S_{\GZ} +  M^2 \int \d^d x \left[ -\left( \overline{\varphi}^a_i
\varphi^a_{i} - \overline{\omega}^a_i \omega^a_i \right)  + 2
\frac{d (N^2 -1)}{\sqrt{2 g^2 N}}  \varsigma \ \gamma^2 \right]
\end{eqnarray}
can be proven analogously as in \cite{Dudal:2008sp}. The vacuum term
$S_{\en}$ will not give rise to any additional counterterms, hence,
\begin{eqnarray}
\frac{\varsigma_0 \gamma_0^2 J_0}{ g_0}   &=& \frac{\varsigma
\gamma^2 J}{ g} \;,
\end{eqnarray}
and consequently, adding this extra term does not give rise to a new
renormalization factor,
\begin{eqnarray}
Z_{\varsigma} &=& Z_g Z_{\gamma^2}^{-1} Z_J^{-1} \;.
\end{eqnarray}

\section{The gluon and ghost propagator}
We shall use the following conventions for the gluon and the ghost
propagator,
\begin{align}
\Braket{A_\mu^a(-p) A_\nu^b (p)}
&=\delta^{ab}\mathcal{D}(p^2)\mathcal{P}_{\mu\nu}(p)\;,
&\Braket{c^a(-p) \overline{c}^b(p)}&=\delta^{ab}\mathcal{G}(p^2)\;,
\end{align}
where $\mathcal{P}_{\mu\nu}(p) =  \delta_{\mu\nu}
-\frac{p_{\mu}p_{\nu}}{p^2}$ is the transverse projector.
\subsection{The gluon propagator}
The tree level gluon propagator corresponding to the action
\eqref{nact} is given by
\begin{equation}\label{gluonprop2}
    \mathcal{D}(p^2)=\frac{p^2 + M^2}{p^4 + M^2p^2 + \lambda^4}\;,
\end{equation}
with
\begin{eqnarray}
\lambda^4 = 2 g^2 N \gamma^4 \;.
\end{eqnarray}
This particular propagator \eqref{gluonprop2} displays the following
properties already at tree level:
\begin{itemize}
\item ${\cal D}(p^2)$ is infrared suppressed due to the presence of the mass
scales $M^2$ and $\lambda^4$.
\item ${\cal D}(0)=\frac{M^2}{\lambda^4}$, i.e. the gluon propagator
does not vanish at zero momentum if $M^2$ is different from zero.
\end{itemize}
These properties seem to be in qualitative accordance with the
lattice data
\cite{Cucchieri:2007md,Bogolubsky:2007ud,Cucchieri:2007rg}. We also
want to stress that the mass term related to
$\overline{\varphi}\varphi-\overline{\omega}\omega$ plays a crucial
role in having ${\cal D}(0)\neq0$, since in the standard
Gribov-Zwanziger scenario, the
gluon propagator necessarily goes to zero.\\

It is instructive to determine the one loop value of the gluon
propagator near zero momentum in order, for example, to produce a
numerical estimate to compare with other methods. Again this
calculation is similar to that performed in $4D$ and we record those
features which are different for our 3-dimensional analysis. First,
we note that for this exercise we will follow \cite{1} where the
form of the gluon and $\varphi$ field $2$-point mixing term is
defined differently. For this intermediate calculation here it is
appropriate to use those conventions and method of \cite{1} since
they have been demonstrated to be consistent with the gluon
suppression and ghost enhancement of the original Gribov analysis at
two loops in $\MSbar$ in $4D$. At the end we have been careful in
converting back to the main conventions of this article in deriving
the one loop freezing value of the gluon propagator. Our method
involves computing the matrix of $2$-point functions comprising the
gluon-$\varphi$ field sector at one loop in the $\MSbar$ scheme and
then inverting this at one loop to determine the quantum corrections
to the tree propagators. However, as we are ultimately only
interested in the zero momentum limit we restrict ourselves to
evaluating the $2$-point functions in the zero momentum limit from
the outset using the vacuum bubble expansion. In this all fourteen
contributing Feynman diagrams are expanded in powers of the external
momentum $p^2$ but truncated at $O((p^2)^2)$. Unlike in $4D$ all the
diagrams are ultraviolet finite and the renormalization of all the
parameters is trivial. In other words one simply replaces bare
quantities by their corresponding renormalized ones. In effect this
is a trivial $\MSbar$ renormalization. In addition the basic one
loop vacuum bubble integral is simple to compute and is given by
differentiating \eqref{pot3} with respect to $m^2$. To determine the
freezing value of the gluon propagator we have adapted the computer
programme used for the 4-dimensional computation,
\cite{Dudal:2008sp}, written in the symbolic manipulation language
{\sc Form}, \cite{2}, to the 3-dimensional case. As indicated
already this is the main reason for adopting the conventions of
\cite{1} here and it requires the replacement of the basic one loop
4-dimensional integrals by their 3-dimensional counterparts in the
{\sc Form} programme. Whilst this may seem a trivial exercise there
is an important aspect to note. In performing automatic Feynman
diagram computations, where the graphs are generated electronically
with the {\sc Qgraf} package, \cite{3}, one has always to ensure the
correctness of the result. Paradoxically an ultraviolet divergent
4-dimensional calculation is easier to check than a finite
3-dimensional one. The reason for this resides in the fact that for
the former the correct non-trivial renormalization constants have to
emerge, and in the case of a gauge theory, these have to satisfy the
Slavnov-Taylor identities. In a finite calculation the luxury of
this check is absent. However, in the present case the adaptation of
a verified programme with minor changes gives us confidence in the
eventual value we will derive. Moreover, a reasonable degree of
consistency with other results, such as lattice methods, adds to our
confidence in the result as will become
evident later.\\

For completeness, the Landau gauge propagators we use are, for an
arbitrary colour group,
\begin{eqnarray}
\langle A^a_\mu(p) A^b_\nu(-p) \rangle &=&
\frac{\delta^{ab}(p^2+M^2)}{[(p^2)^2+M^2 p^2+C_A\gamma^4]}
P_{\mu\nu}(p)\;,
\nonumber \\
\langle A^a_\mu(p) \overline{\varphi}^{bc}_\nu(-p) \rangle &=& -~
\frac{f^{abc}\gamma^2}{\sqrt{2}[(p^2)^2+M^2 p^2+C_A\gamma^4]}
P_{\mu\nu}(p)\;,
\nonumber \\
\langle \varphi^{ab}_\mu(p) \overline{\varphi}^{cd}_\nu(-p) \rangle
&=& -~ \frac{\delta^{ac}\delta^{bd}}{(p^2+M^2)}\eta_{\mu\nu} ~+~
\frac{f^{abe}f^{cde}\gamma^4}{(p^2+M^2)[(p^2)^2+M^2
p^2+C_A\gamma^4]} P_{\mu\nu}(p)\;,
\end{eqnarray}
where $f^{abc}$ are the colour group structure constants and the
appearance of $1/\sqrt{2}$ derives from the conventions of
\cite{1}. We formally define the matrix of one loop corrections to
the $2$-point functions as
\begin{eqnarray}\label{1PImatrix}
&& \left(
\begin{array}{cc}
p^2 \delta^{ac} & - \gamma^2 f^{acd} \\
- \gamma^2 f^{cab} & - (p^2+M^2) \delta^{ac} \delta^{bd} \\
\end{array}
\right) + \left(\begin{array}{cc}
X \delta^{ac} & U f^{acd} \\
N f^{cab} & Q \delta^{ac} \delta^{bd} + W f^{ace} f^{bde} + R
f^{abe} f^{cde}
+ S d_A^{abcd} \\
\end{array}
\right) g^2 ~+~ O(g^4)\;,
\end{eqnarray}
in the $\left\{\! \frac{1}{\sqrt{2}} A^a_\mu, \varphi^{ab}_\mu
\right\}$ basis where the common tensor $P_{\mu\nu}(p)$ has been
removed, $d_A^{abcd}$ is defined by, \cite{4},
\begin{equation}
d_A^{abcd} ~=~ \frac{1}{6} \mbox{Tr} \left( T_A^a T_A^{(b} T_A^c
T_A^{d)} \right)\;
\end{equation}
and $(T_A^a)_{bc}$~$=$~$-$~$if^{abc}$ is the adjoint representation
of the color group generators. Given the set of formal one loop
$2$-point functions, it is easy to invert the matrix to one loop and
formally derive the corresponding matrix of propagators, as
\begin{eqnarray}\label{matrix}
&& \left(
\begin{array}{cc}
\frac{(p^2+M^2)}{[(p^2)^2+M^2p^2+C_A\gamma^4]} \delta^{cp} &
- \frac{\gamma^2}{[(p^2)^2+M^2p^2+C_A\gamma^4]} f^{cpq} \\
- \frac{\gamma^2}{[(p^2)^2+M^2p^2+C_A\gamma^4]} f^{pcd} & -
\frac{1}{(p^2+M^2)} \delta^{cp} \delta^{dq}
+ \frac{\gamma^4}{(p^2+M^2)[(p^2)^2+M^2p^2+C_A\gamma^4]} f^{cdr} f^{pqr} \\
\end{array}
\right) \nonumber \\
&&+ \left(
\begin{array}{cc}
A \delta^{cp} & C f^{cpq} \\
E f^{pcd} & G \delta^{cp} \delta^{dq} + J f^{cpe} f^{dqe} + K
f^{cde} f^{pqe}
+ L d_A^{cdpq} \\
\end{array}
\right) g^2 ~+~ O(g^4) ~.
\end{eqnarray}
We refer to Section 3 in \cite{1} for a detailed discussion
about the color structures emerging in this matrix.

As we are specifically interested in the gluon propagator the only
quantity of importance here is $A$ and it is formally the same as in
\cite{Dudal:2008sp}. In other words,
\begin{eqnarray}
\hspace{-2mm}A &=& -~ \frac{1}{[(p^2)^2+M^2p^2+C_A\gamma^4]^2}
\times \left[ (p^2+M^2)^2 X - C_A \gamma^2 (N+U) (p^2+M^2) + C_A
\gamma^4 \left( Q + C_A R + \half C_A W \right) \right] \, .
\end{eqnarray}
This concludes the derivation of the formal aspects of the
computation of the one loop propagator corrections. The actual
values of the $2$-point function
contributions now need to be inserted from the vacuum bubble expansion.\\

Let us point out here that we are looking at the connected
gluon two-point function, which is the relevant quantity, also
measured on the lattice. The reader might notice that the lowest
order part of the gluon propagator in \eqref{matrix} can also be
obtained by integrating out the additional
$(\overline{\varphi},\varphi)$- fields in the quadratic
approximation and construct the tree level gluon propagator in this
fashion. In the case $M^2=0$, this also corresponds with the result
for the gluon propagator from the original semiclassical approach by
Gribov \cite{Gribov:1977wm}. It is worth mentioning that this
particular kind of propagator has been used frequently as an ansatz
for a long time to do lattice fits,
see e.g. \cite{Cucchieri:2004mf}.\\

A key difference in evaluating the relevant fourteen Feynman
diagrams derives from the dimensionality of the basic
$d$-dimensional integral defined by
\begin{equation}
I(\mu) ~=~ \int \frac{\d^d k}{(2\pi)^d} \frac{1}{(k^2 + \mu^2)}
\end{equation}
for a generic mass $\mu$. On dimensional grounds $I(\mu)$ has mass
dimensions of $(d-2)$. In four dimensions this of course corresponds
to a mass dimension of two. By contrast in three dimensions this
drops to one. However, in each of the tree propagators there is a
denominator factor which is quadratic in $p^2$ which gives two
roots. In $4D$ this leads to a very simple partial fraction into two
terms and hence the simple evaluation of the basic vacuum integral
$I(\mu)$. In $3D$ the situation is more complicated. The same
partial fraction emerges but one has to choose the sign of the
square root of the mass term appearing as $\mu^2$ for the evaluation
of the integral in terms of an object of mass dimension one. In
principle this could lead to several analyses depending on the
choice of sign in the square root. Another way of viewing this is to
realize that in three dimensions one has to consider the common
propagator factor as quartic in $\sqrt{p^2}$ and find four roots.
For completeness these are
\begin{eqnarray}
m^{\pm}_+ &=& \pm \frac{1}{2} \left[ \sqrt{M^2 + 2 \sqrt{C_A}
\gamma^2}
+ \sqrt{M^2 - 2 \sqrt{C_A} \gamma^2} \right] \nonumber \\
m^{\pm}_- &=& \pm \frac{1}{2} \left[ \sqrt{M^2 + 2 \sqrt{C_A}
\gamma^2} - \sqrt{M^2 - 2 \sqrt{C_A} \gamma^2} \right] ~.
\label{massroots}
\end{eqnarray}
In the limit where $M$~$\rightarrow$~$0$ one recovers the four more
recognizable poles of the usual Gribov propagator
\begin{equation}
\lim_{M\rightarrow 0} m^{\pm}_+ ~=~ \pm \frac{1}{2} (1+ i)\gamma
~~,~~ \lim_{M\rightarrow 0} m^{\pm}_- ~=~ \pm \frac{1}{2} (1 -
i)\gamma ~.
\end{equation}
To resolve which is the correct choice of signs for the integral, we
recall that the gap equation for the Gribov mass derives from a one
loop calculation which also involves the basic integral $I(\mu)$.
Examining that calculation in order to have a non-trivial solution
with a {\em positive} coupling constant, one has to take $m^+_+$ and
$m^+_-$. Interestingly in our calculation the only square root which
remains in the leading order vacuum bubble expansion relevant for
the gluon propagator freezing is $\sqrt{M^2 + 2 \sqrt{C_A}
\gamma^2}$ which is always real. So there are no issues concerning
the relative sizes of $M^2$ and $\gamma^2$ which could have led to a
complex value in the square roots of (\ref{massroots}). Given these
considerations and being careful to revert to our general
conventions, we finally find,
\begin{eqnarray}\label{dnul}
\mathcal{D}(0) &=& \frac{M^2}{\lambda^4}+ \frac{g^2 N}{4 \pi} \frac{M^4}{\lambda^8} \left[\frac{ M^4}{\lambda^4}  \sqrt{M^2 + 2 \lambda^2}  -  \frac{M^5}{\lambda^4}  -  \frac{M^2}{\lambda^2} \sqrt{M^2 + 2 \lambda^2} -\frac{17}{12} M^2 \lambda^2 \frac{\sqrt{M^2 + 2 \lambda^2}}{M^4 - 4 \lambda^4} + \frac{13}{4} \lambda^4 \frac{\sqrt{M^2 + 2 \lambda^2}}{M^4 - 4 \lambda^4} \right. \nonumber\\
&& \left.  -\frac{5}{3}\lambda^6 M^2 \frac{\sqrt{M^2 + 2
\lambda^2}}{(M^4-4 \lambda^4)^2}+ \frac{10}{3}\lambda^8
\frac{\sqrt{M^2 + 2 \lambda^2}}{(M^4 - 4 \lambda^4)^2} + \frac{7}{4}
M - \frac{1}{4}\sqrt{M^2 + 2 \lambda^2} \right] \;.
\end{eqnarray}
From the previous expression, we can still conclude that for a non
zero value of $M^2$, $\mathcal D(0) \not= 0$.

\subsection{The ghost propagator}
We have found that the one loop corrected ghost propagator can be
written as
\begin{eqnarray}\label{ghost}
\mathcal{G}(k^2) &=&  \frac{1}{k^2} \frac{1}{1- \sigma(k^2)}\;,
\end{eqnarray}
where $\sigma(k^2)$ is the following momentum dependent function
\begin{eqnarray}\label{sigmak}
\sigma(k^2) &=& g^2N \frac{k_{\mu} k_{\nu}}{k^2} \int \frac{\d^3
q}{(2\pi)^3} \frac{1}{(k-q)^2} \frac{q^2 + M^2}{q^4 + M^2q^2 +
\lambda^4}\mathcal{P}_{\mu\nu}(q) \;.
\end{eqnarray}
Calculating this integral explicitly, we find
\begin{eqnarray}\label{sigmakuitgerekend}
\sigma(k^2) &=& \frac{N g^2}{32 k^3 \pi} \Biggl\{ \frac{1}{M^2- \sqrt{M^4 - 4 \lambda^4}} \left( 1 + \frac{M^2}{\sqrt{M^4 - 4 \lambda^4}} \right) \Biggl[ \sqrt{2} k^3 \sqrt{M^2 - \sqrt{M^4 - 4 \lambda^4}} - \frac{k}{\sqrt{2}} \left( M^2 - \sqrt{M^4 - 4 \lambda^4}\right)^{3/2} \nonumber\\
&& \hspace{4cm} - k^4 \pi + \frac{1}{2} \left(2 k^2 + M^2 - \sqrt{M^4 - 4 \lambda^4} \right)^2 \arctan \frac{\sqrt{2} k}{ \sqrt{M^2 - \sqrt{M^4 - 4 \lambda^4}}} \Biggr]\nonumber\\
 && \hspace{0.9cm}+ \frac{1}{M^2 + \sqrt{M^4 - 4 \lambda^4}}  \left( 1 - \frac{M^2}{\sqrt{M^4 - 4 \lambda^4}} \right) \Biggl[ \sqrt{2} k^3 \sqrt{M^2 + \sqrt{M^4 - 4 \lambda^4}} - \frac{k}{\sqrt{2}} \left( M^2 + \sqrt{M^4 - 4 \lambda^4}\right)^{3/2} \nonumber\\
&& \hspace{4cm} - k^4 \pi + \frac{1}{2} \left(2 k^2 + M^2 +
\sqrt{M^4 - 4 \lambda^4} \right)^2 \arctan \frac{\sqrt{2} k}{
\sqrt{M^2 + \sqrt{M^4 - 4 \lambda^4}}} \Biggr]\Biggr\} \;.
\end{eqnarray}
In order to find the behavior of the ghost propagator near zero
momentum we take the limit $k^2 \to 0$ in equation \eqref{sigmak},
\begin{eqnarray}\label{gg}
\sigma(0) &=& g^2N
\frac{2}{3}\int\frac{\d^3q}{(2\pi)^3}\frac{1}{q^2}\frac{q^2 +
M^2}{q^4+ M^2 q^2 + \lambda^4} ~=~  \frac{ g^2N}{6\pi} \frac{M^2 +
\lambda^2}{\lambda^2 \sqrt{M^2 + 2 \lambda^2}}\;,
\end{eqnarray}
which can of course be obtained by taking the limit $k^2 \to 0$ in
expression \eqref{sigmakuitgerekend}. Similarly, one
can check that $\sigma\to0$ for $k^2\to\infty$ and/or $M^2\to\infty$.\\

Before drawing any conclusions, we still need to have a look at the
gap equations, which shall fix $\lambda^2$ as a function of $M^2$.

\subsection{The gap equations}
We begin with the first gap equation \eqref{horizoncond2} in order
to express $\lambda$ as a function of $M^2$. The effective action at
one loop order is given by
\begin{eqnarray}
\Gamma_\gamma^{(1)} &=& -d(N^{2}-1)\gamma^{4} + 2\frac{d (N^2
-1)}{\sqrt{2 g^2 N}}  \varsigma \ \gamma^2 M^2
+\frac{(N^{2}-1)}{2}\left( d-1\right) \int \frac{\d^{d}q}{\left(
2\pi \right) ^{d}} \ln \frac{q^4 + M^2 q^2 + 2 g^2 N \gamma^2}{ q^2
+ M^2} \;.
\end{eqnarray}
With $\lambda^4 = 2 g^2 N \gamma^4$, we rewrite the previous
expression,
\begin{eqnarray}
\mathcal{E}^{(1)} &=&  \frac{\Gamma_\gamma^{(1)}}{N^2 - 1} \frac{2
g^2 N}{d} ~=~ - \lambda^4  + 2 \varsigma \lambda^2 M^2 + g^2 N
\frac{ d-1}{d} \int \frac{\d^{d}q}{\left( 2\pi \right) ^{d}} \ln
\frac{q^4 + M^2 q^2 + \lambda^4}{ q^2 + M^2} \;,
\end{eqnarray}
and apply the gap equation \eqref{horizoncond2},
\begin{eqnarray}\label{gapintegraal}
0&=&  -1 + \varsigma \frac{M^2}{\lambda^2} + g^2 N \frac{d-1}{d}
\int \frac{\d^{d}q}{\left(2\pi \right) ^{d}}  \frac{1}{q^4 + M^2 q^2
+ \lambda^4} \;.
\end{eqnarray}
In 3 dimensions, the integral in this gap equation is finite,
resulting in,
\begin{eqnarray}\label{gapuitgerekend}
0&=&  -1 + \varsigma \frac{M^2}{\lambda^2} + \frac{g^2 N}{6\pi}
\frac{1}{\sqrt{M^2 + 2 \lambda^2}} \;.
\end{eqnarray}
This expression will fix $\lambda^2$ as a function of $M^2$,
i.e.~$\lambda^2(M^2)$ once we have found an explicit value for $\varsigma$.\\

This explicit value for $\varsigma$ will be provided by the second
gap equation \eqref{bound}. From expression \eqref{sigmak}, one
finds
\begin{eqnarray}
\sigma(0) &=& g^2N \frac{d- 1}{d}\int\frac{\d^dq}{(2\pi)^d}\frac{1}{q^2}\frac{q^2 + M^2}{q^4+ M^2 q^2 + \lambda^4} \nonumber\\
   &=& g^2N  \frac{d-1}{d} \int\frac{\d^dq}{(2\pi)^d}\frac{1}{q^4+ M^2 q^2 + \lambda^4} + M^2 g^2N \frac{d-1}{d}\int\frac{\d^dq}{(2\pi)^d}\frac{1}{q^2}\frac{1}{q^4+ M^2 q^2 + \lambda^4}\;.
\end{eqnarray}
Therefore, we can rewrite the first gap equation
\eqref{gapintegraal} as
\begin{equation}\label{sigmazero}
    0~=~\sigma(0) - 1-M^2\frac{d-1}{d}g^2N\int \frac{\d^dq}{(2\pi)^d}\frac{1}{q^2}\frac{1}{q^4+M^2q^2+\lambda^4}+\varsigma\frac{M^2}{\lambda^2} \;.
\end{equation}
The second gap equation can then subsequently be obtained by acting
with $\frac{\p}{\p M^2}$ on the previous expression and setting
$M^2=0$. Doing so, we find
\begin{equation}\label{simpel}
    -\frac{d-1}{d}g^2N\int \frac{\d^dq}{(2\pi)^d}\frac{1}{q^2}\frac{1}{q^4+\lambda^4(0)}+\varsigma\frac{1}{\lambda^2(0)}~=~0
\end{equation}
by keeping \eqref{bound} in mind. Setting $M^2 = 0$ in
\eqref{gapuitgerekend} yields,
\begin{eqnarray}
\sqrt{2\lambda^2(0)} &=&   \frac{g^2 N}{6 \pi}\;.
\end{eqnarray}
Proceeding with equation \eqref{simpel}, we find the following
simple solution for $\varsigma$,
\begin{equation}\label{varsigma}
    \varsigma~=~\frac{1}{12\pi}\frac{2}{\sqrt{2}}\frac{g^2N}{\lambda(0)}=1\;.
\end{equation}
In summary, the following expression,
\begin{eqnarray}\label{gapfinal}
 \frac{g^2 N}{6\pi} \frac{1}{\sqrt{M^2 + 2 \lambda^2}} &=&   1 -  \frac{M^2}{\lambda^2}
\end{eqnarray}
fixes $\lambda^2(M^2)$.

\subsection{The ghost propagator at zero momentum}
At this point, we have all the information we need to take a closer
look at the ghost propagator at zero momentum. From equation
\eqref{gg} and \eqref{gapfinal}, we find
\begin{eqnarray}
    \sigma(0) &=& \left( \frac{M^2}{\lambda^2} + 1 \right)  \left(  1 -  \frac{M^2}{\lambda^2} \right) ~=~ 1 -  \frac{M^4}{\lambda^4}\;.
\end{eqnarray}
From this expression we can make several observations. Firstly, when
$M^2 =0$, we find that $\sigma(0) = 1$, which is exactly the result
obtained in the original Gribov-Zwanziger action
\cite{Gribov:1977wm,Zwanziger:1992qr}. Consequently, the ghost
propagator \eqref{ghost} is enhanced and behaves like $1/k^4$ in the
low momentum region. Secondly, for any $M^2 >0$, $\sigma(0)$ is
smaller than 1. By contrast, without the inclusion of the extra
vacuum term, $\sigma(0)$ would always be bigger than 1, which can be
observed from expression \eqref{sigmazero}. Therefore, it is
absolutely necessary to include this term. With $\sigma(0)$ smaller
than 1, the ghost propagator is not enhanced and behaves as $1/k^2$.

\section{Study of the dynamical effects related to $M^2$}
\subsection{Variational perturbation theory}
In this section, we shall rely on variational perturbation theory in
order to find a dynamical value for the hitherto arbitrary mass
parameter $M^2$. Specifically, we follow
\cite{Dudal:2008sp,Jackiw:1995nf} and introduce $M^2$ as a
variational parameter into the theory by replacing the action
\eqref{totaleactie} by
\begin{equation}\label{var}
    S_{GZ} +   (1-\ell^k)M^2 \int \d^d x \left[ - \left(\overline{\varphi}^a_i \varphi^a_{i} - \overline{\omega}^a_i \omega^a_i\right)   + 2 \frac{d (N^2 -1)}{\sqrt{2 g^2 N}}  \varsigma \ \gamma^2
    \right]\;,
\end{equation}
where $\ell$ serves as the loop counting parameter, and formally
$\ell=1$ at the end \cite{Dudal:2008sp}. In this fashion, it is
clear that the original starting action $S_{\GZ}$ has not been
changed. We have in fact added the terms in $M^2$, and subtracted
them again at $k$ orders higher in the loop expansion. We shall set
$k = 2$ as we are working up to one loop. Taking $k=1$ would destroy
the effect of the vacuum term $S_{\en}$, which would be inconsistent
as
explained before.\\

We shall discuss a possible option to fix $M^2$. For this, we can
expand any quantity $\mathcal Q^{(n)}$, which is evaluated to a
certain order $n$, in powers of $\ell$, cut this series to the order
$n$ and set $\ell=1$. We can then require that $\frac{\p \cal Q}{\p
M^2 }=0$, i.e.~ the principle of \textit{minimal sensitivity}
\cite{Stevenson:1981vj}. This latter requirement can be well
motivated since an exact calculation of $\cal Q$ using the action
\eqref{var}, taking the full nonperturbative nature of the theory
into account, would lead to an $M^2$-independent result, since
$\ell=1$ after all. At a finite order in $\ell$ however, some
residual $M^2$ dependence will enter the result. In this way, we can
hope to capture some relevant nontrivial information yet at finite
order, encoded in the parameter $M^2$. We then mimic the
independence on $M^2$ of the exact result just by imposing $\frac{\p
\cal Q}{\p M^2 }=0$ and thereby fixing $M^2$. If, however, no
optimal value for $M^2$ is found, one can impose that $\frac{\p^2
\cal Q}{ (\p M^2)^2 }=0$ \cite{Stevenson:1981vj}.

\subsection{The ghost propagator}
In order to obtain a dynamical value for the ghost propagator in the
presence of $M^2$, we shall now utilize the variational procedure
outlined in the previous subsection. We start with the expression
\eqref{sigmak} for $\sigma(k^2)$ and replace $g^2$ with $\ell g^2$
and $M^2$ with $(1-\ell^2)M^2$, expand up to order $\ell$ and set
$\ell=1$. Doing so, we recover again the same expression
\eqref{sigmak}. Following an analogous procedure for the gap
equations, also results in the same expression \eqref{gapfinal}.
This latter equation determines $\lambda^2(M^2)$, which can be
plugged in the expression of $\sigma(k^2)$, thereby making
$\sigma(k^2)$ only a
function of $M^2$, next to the momentum dependence. \\

\begin{figure}[t]
  \centering
      \includegraphics[width=7cm]{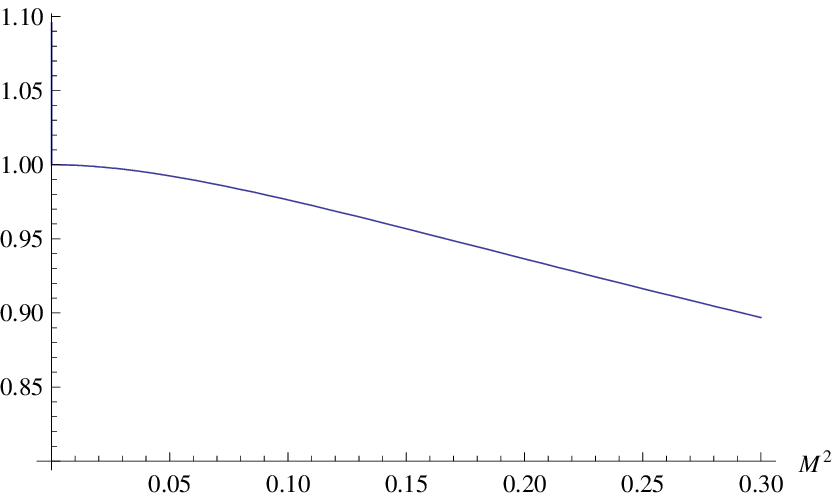}
     \includegraphics[width=7cm]{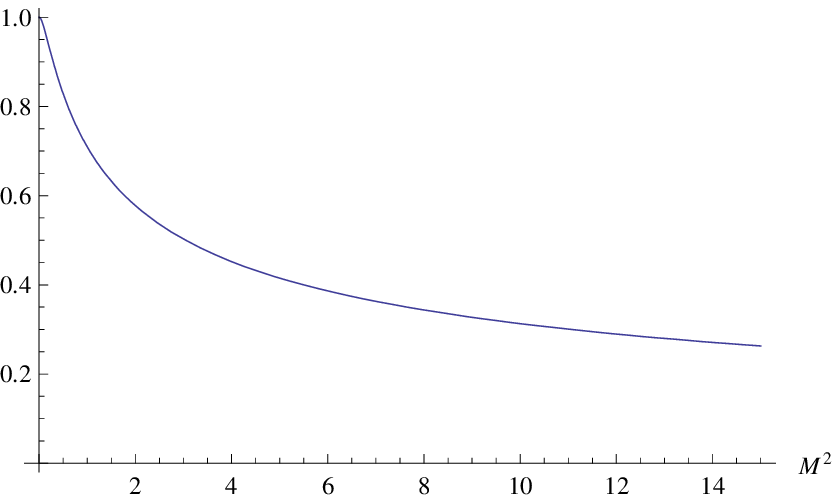}
  \caption{$\sigma(0)$ in function of $M^2$ in units $\frac{g^2 N}{6 \pi} =1$.}  \label{fig1}
\end{figure}

Firstly, let us investigate $\sigma(k^2)$ at zero momentum, which is
the key-point of this paper. FIG.~\ref{fig1} displays $\sigma(0)$ as
a function of $M^2$. We observe that $\sigma(0)$ is indeed smaller
than 1 for all $M^2 > 0$ as already shown analytically in the
previous section. We also find a smooth limit of $\sigma(0)$ for
$M^2 \to 0$ required by the second gap equation \eqref{bound}, as
can be seen from the left figure. According to the principle of
minimal sensitivity, we have to search for an extremum,
i.e.~$\frac{\p  \sigma(0)}{\p M^2 }=0$. Unfortunately, there is no
such an extremum present.
 Nevertheless, we do find a point of inflection,
$\frac{\p^2 \sigma(0)}{ (\p M^2)^2 }=0$ at $M^2 = 0.185
\left(\frac{g^2N}{6\pi}\right)^2$. Taking this value for $M^2$, we
find
\begin{eqnarray}
\sigma(0) &=&  0.94\;,
\end{eqnarray}
which is indeed smaller than one and results in a non-enhanced
behavior of the ghost propagator. This value needs to be
compared with the lattice value of $\sigma(0) = 0.79$, which can be extracted from the data in \cite{Cucchieri:2008fc}. \\

\begin{figure}[t]
  \centering
      \includegraphics[width=10cm]{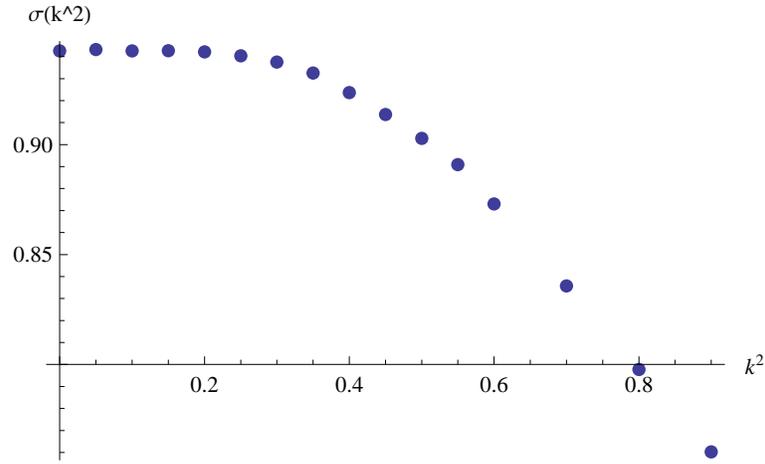}
  \caption{The optimal $\sigma(k^2)$ in function of $k^2$ in units of $\frac{g^2 N}{6 \pi }=1$.}  \label{fig2}
\end{figure}

Secondly, let us have a look at this point of inflection, when ``turning on'' the momentum $k^2$. As shown in TABLE \ref{table1}, we observe that $M^2$ will decrease, until it will vanish at $k^2
\approx 0.55$. The corresponding $\sigma(k^2)$ is displayed in FIG.~\ref{fig2}. We thus see that we find some kind of a momentum dependent effective mass $M^2(k^2)$, which disappears when $k$ grows. This could have been anticipated, as we naturally expect that the deep ultraviolet sector should be hardly affected. Let us also notice here that $\sigma(k^2)$ is a decreasing function, as can be explicitly checked from FIG.~\ref{fig2}. This of course means that we are staying within the horizon for any value of the momentum.\\
\begin{table}[t]
\renewcommand{\arraystretch}{2}
\begin{center}
\begin{tabular}{|c||cccccccccccc|}
        \hline
        $k^2 $& 0 & 0.05 & 0.1 & 0.15 & 0.2 & 0.25 & 0.30 & 0.35 & 0.40 & 0.45 &0.50 & 0.55  \\
        \hline
        $M_{\mathrm{min}}^2$& 0.19 & 0.16 & 0.14 & 0.12 & 0.10 & 0.08 & 0.06 & 0.04 & 0.03 & 0.02 & 0.01 & 0 \\
        \hline
\end{tabular}
\end{center}
\caption{Some $M^2_{\min}$ for different $k^2$ in units $\frac{g^2
N}{6 \pi}=1$.}\label{table1}
\end{table}

\begin{figure}[t]
  \centering
      \includegraphics[width=10cm]{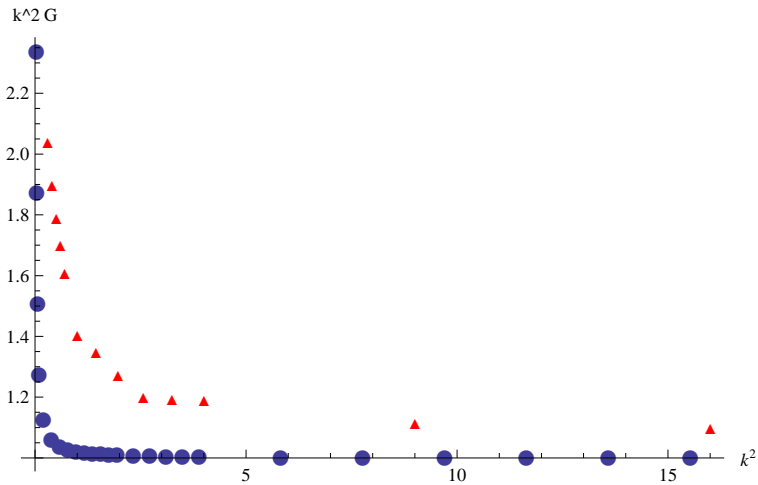}
  \caption{The optimal $k^2 \mathcal{G}(k^2)$ in function of $k^2$ in units of GeV. The lattice (our analytical) results are indicated with triangles (dots). The error bars on the
  lattice data are roughly of the size of the triangles. }  \label{fig3}
\end{figure}

\begin{figure}[t]
  \centering
      \includegraphics[width=10cm]{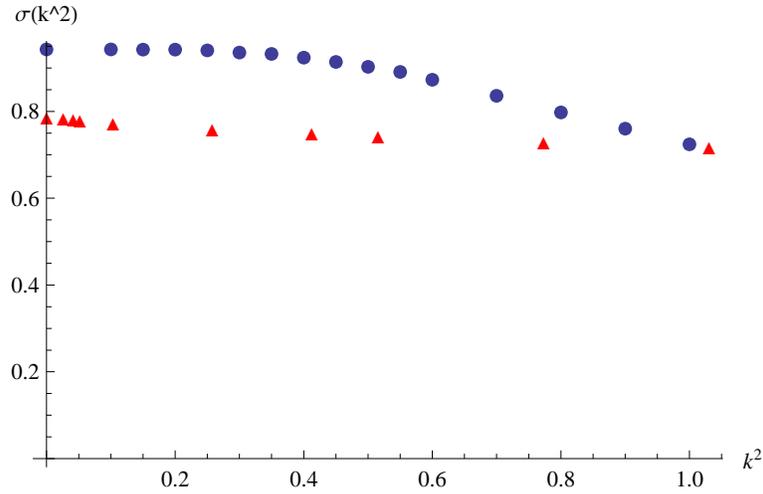}
  \caption{The optimal $\sigma(k^2)$ in function of $k^2$ in units of $\frac{g^2 N}{6 \pi }=1$. The lattice (our analytical) results are indicated with triangles (dots).}  \label{fig3accent}
\end{figure}

To compare our results with available lattice data, we
must make the conversion to physical units of GeV. In
\cite{Lucini:2002wg} a continuum extrapolated value for the ratio
$\sqrt{\sigma}/g^2$ was given for several gauge groups; in
particular, $\sqrt{\sigma}/g^2 \approx0.3351$ for $SU(2)$. Further,
$\sqrt{\sigma}$ stands for the square root of the string tension.
For this quantity, we used the input value of $\sqrt{\sigma} = 0.44$
GeV as in \cite{Cucchieri:2004mf}. Therefore, for $SU(2)$, we find,
\begin{eqnarray}
\left(\frac{g^2 N}{6 \pi}\right)^2 & \approx & 0.0194 \mathrm{GeV}^2\;.
\end{eqnarray}
In FIG.~\ref{fig3}, we have plotted the lattice as well
our analytical result for the ghost dressing function,
$k^2\mathcal{G}(k^2)$, in units of GeV. We used the numerical data
of \cite{Cucchieri:2008fc,Cucchieri:2008qm}, adapted to our needs.
We observe that for sufficiently large $k^2$, the lattice data and
our analytical results converge. In this case, the novel mass $M^2$
becomes zero as advocated earlier, meaning that we are back in the
usual Gribov-Zwanziger scenario. For smaller $k^2$, we found it more
instructive to compare the lattice estimate of $\sigma(k^2)$ with
our value as the errors on $k^2\mathcal{G}(k^2) = \frac{1}{1-
\sigma(k^2)}$ are becoming large when looking at
 $\sigma(k^2)$
close to 1. FIG.~\ref{fig3accent} displays $\sigma(k^2)$ in units of
$\frac{g^2 N}{6 \pi }=1$ up to $k^2 = 1 \times \left(\frac{g^2 N}{6
\pi }\right)^2 = 0.0194\ \mathrm{GeV}^2$. We see that both results
are in reasonable agreement, especially if we keep in mind that we
have only calculated $\sigma(k^2)$ in a first order approximation.

\subsection{The gluon propagator}
\begin{figure}[h]
  \centering
      \includegraphics[width=10cm]{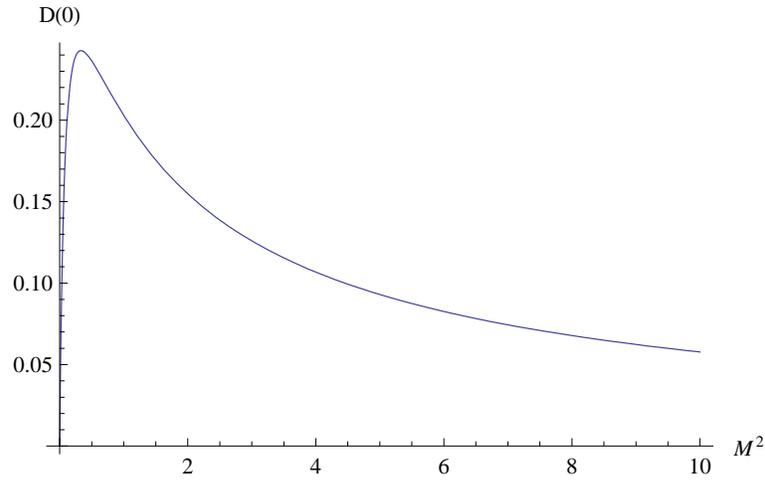}
  \caption{$\mathcal D(0)$ in units $\frac{g^2N}{6\pi}=1$.} \label{fig4}
\end{figure}

We can apply an analogous procedure for the gluon propagator. This
propagator at zero momentum, $\mathcal{D}(0)$ is displayed in
FIG.~\ref{fig4}. We immediately see that there is an extremum at
$M^2 = 0.33\ \left(\frac{g^2 N}{6 \pi}\right)^2$, resulting in
\begin{eqnarray}
\mathcal{D}(0) &=& \frac{0.24}{  \left(\frac{g^2 N}{6
\pi}\right)^2}.
\end{eqnarray}
Doing the conversion to physical units again, we find for $SU(2)$,
\begin{eqnarray} \label{eindres}
\mathcal{D}(0) &\approx&  \frac{12}{  \mathrm{GeV}^2}\,.
\end{eqnarray}
This value can be compared with the bounds derived from
a partially numerical and partially analytical derivation
\cite{Cucchieri:2007rg}:
\begin{eqnarray}
  \frac{1.2}{  \mathrm{GeV}^2} < \mathcal{D}(0) <  \frac{12}{
  \mathrm{GeV}^2}\;.
\end{eqnarray}
We recall that our value $\mathcal{D}(0) = 12 /\mathrm{GeV}^2$ is a
first order approximation and is only of qualitative nature.
Nevertheless, this value is still consistent with the
boundaries of the lattice data set in \cite{Cucchieri:2007rg}.

\subsection{Violation of positivity}
\begin{figure}[t]
  \begin{center}
    \subfigure[\;$t=1$]{\includegraphics[width=5cm]{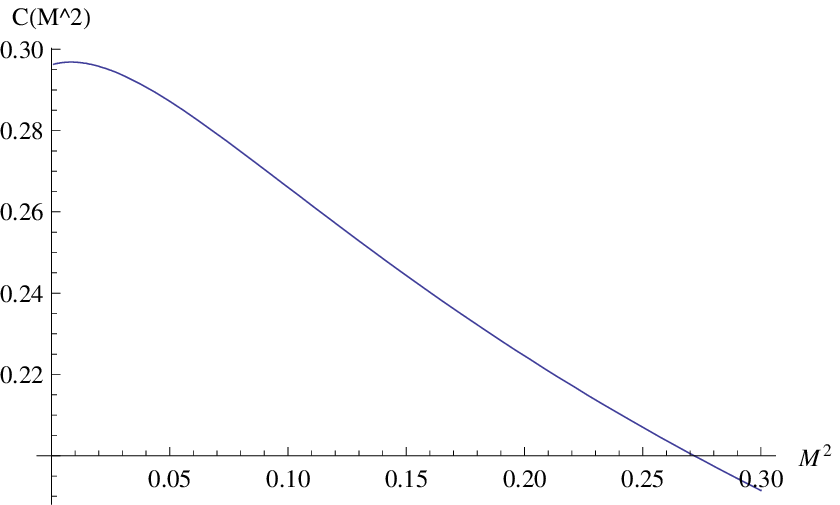} \label{fig1a}}
    \hspace{0.5cm}
    \subfigure[\;$t=1.5$]{\includegraphics[width=5cm]{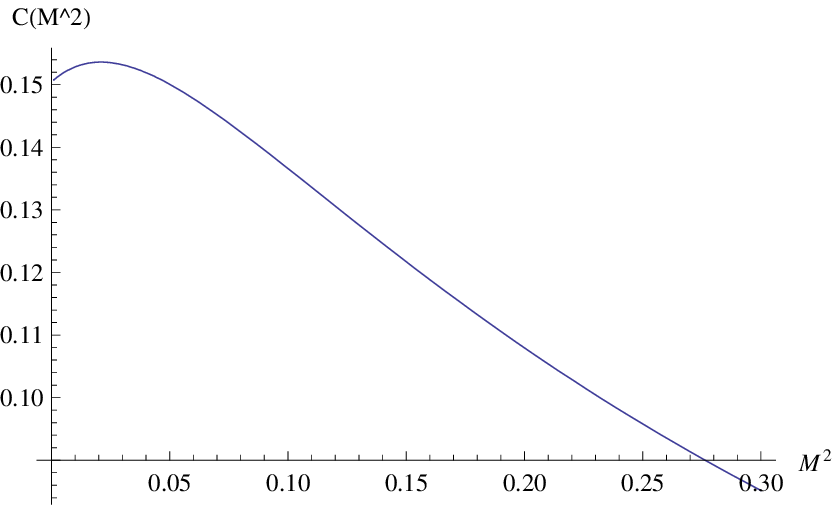}\label{fig1b}}
    \hspace{0.5cm}
    \subfigure[\;$t=2$]{\includegraphics[width=5cm]{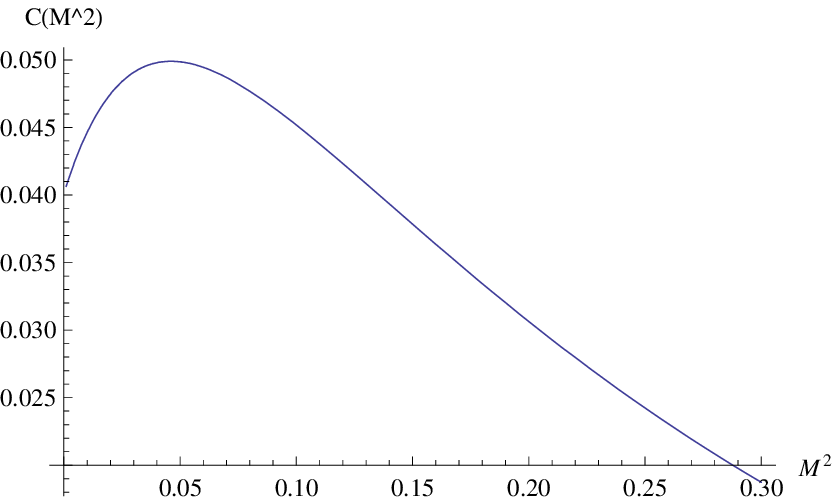}\label{fig1c}}
    \\
    \subfigure[\;$t=2.5$]{\includegraphics[width=5cm]{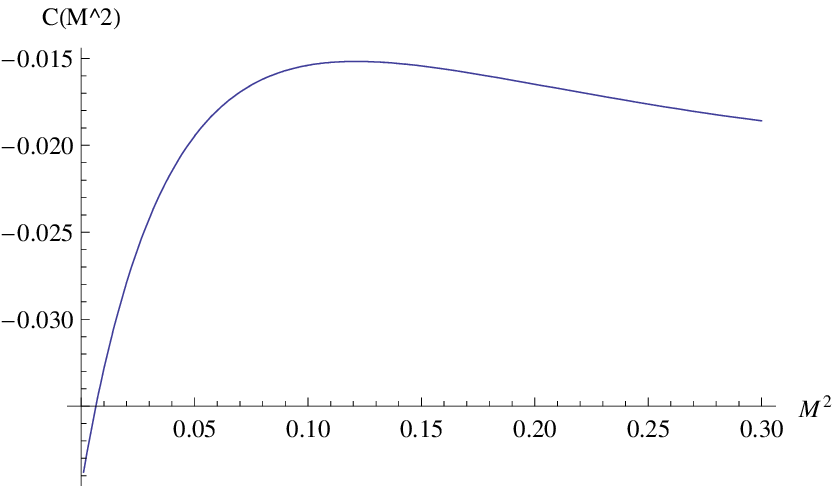}\label{fig1d}}
     \hspace{0.5cm}
    \subfigure[\;$t=3$]{\includegraphics[width=5cm]{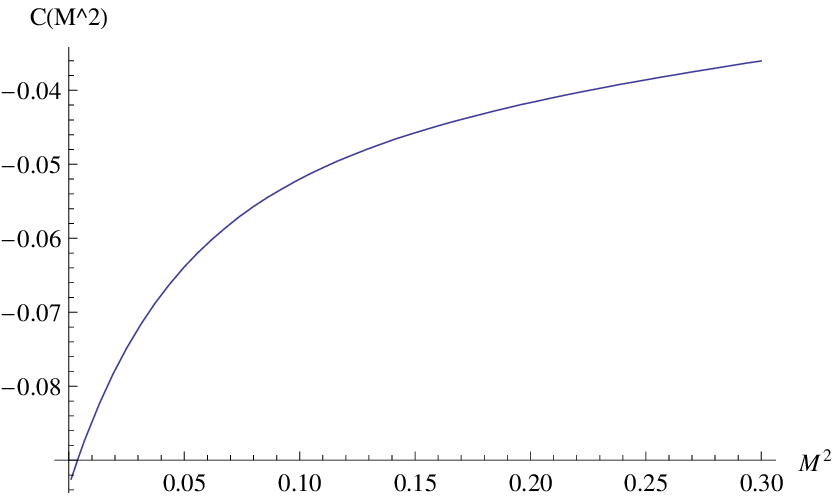}\label{fig1c}}
  \end{center}
  \caption{$\mathcal{C}(t,M^2)$ for a few values of $t$ in function of $M^2$, in units of fm.} \label{fig5}
\end{figure}

We shall investigate if a gluon propagator of the type
\eqref{gluonprop2} displays a violation of positivity, another fact
which is reported by the lattice data \cite{Cucchieri:2004mf}.
Following the analysis of \cite{Cucchieri:2004mf}, the Euclidean
gluon propagator can be expressed by a K\"{a}llen-Lehmann
representation as
\begin{equation}\label{spectr1}
    {\cal D}(p^2)=\int_0^{+\infty} \d M^2\frac{\rho(M^2)}{p^2+M^2}\;,
\end{equation}
whereby the spectral density $\rho(M^2)$ should be positive to make
possible the interpretation of the fields in term of stable
particles. One can define the temporal correlator
\cite{Cucchieri:2004mf}
\begin{equation}\label{spectr2}
    \mathcal{C}(t)=\int_0^{+\infty}\d M\rho(M^2)e^{-Mt}\;,
\end{equation}
which is certainly positive for positive $\rho(M^2)$. However, if
$\mathcal{C}(t)$ becomes negative for certain $t$, $\rho(M^2)$
cannot be positive for all $M^2$, indicating that the gluon is not a
stable physical excitation. Since $\mathcal{C}(t)$ can be rewritten
as
\begin{equation}\label{spectr3}
    \mathcal{C}(t)=\frac{1}{2\pi}\int_{-\infty}^{+\infty} \e^{-ipt}{\cal D}(p^2)\d p\;.
\end{equation}
we must in fact calculate the $1D$ Fourier transformation of ${\cal
D}(p^2)$. In FIG.~\ref{fig5} the Fourier transforms,
$\mathcal C (t, M^2)$ are shown for different $t$ in units of fm.\\

\begin{figure}[h]
  \centering
      \includegraphics[width=7cm]{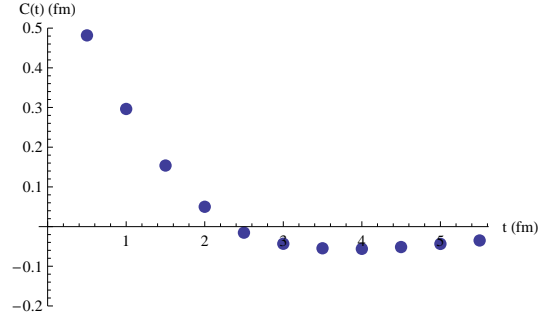}
  \caption{$\mathcal{C}(t)$ in terms of $t$ in units of fm in the refined GZ case.} \label{fig6}
\end{figure}

To determine $M^2$, we again rely on the variational setup. We
observe that for small $t$, there is no real extremum. However, for
a certain $t\sim 1$, an extremum emerges at $M^2\sim 0$. This
extremum starts to grow for increasing $t$, but at the same time,
the curve flattens out. Therefore, starting from $t\sim 2.6 $, the
extremum disappears again. Hence, we have
taken\footnote{$M^2\sim0.18$ is the maximal extremal value,
corresponding to $t\sim2.6$.} $M^2=0$ for $t<1$, and $M^2=0.18$ for
$t>2.6$. The resulting temporal correlator is displayed in
FIG.~\ref{fig6}.
We clearly observe a violation of positivity, which is in agreement
with the lattice data (see FIG.~4 of \cite{Cucchieri:2004mf}).
Although this agreement is only at a qualitative level, the shape of
$\mathcal{C}(t)$ is
very similar. \\

\begin{figure}[t]
  \centering
      \includegraphics[width=7cm]{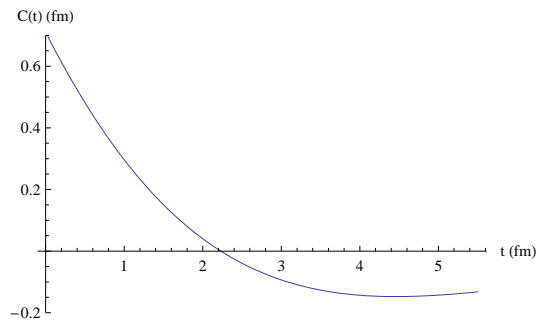}
  \caption{$\mathcal{C}(t)$ in terms of $t$ in units of fm in the pure GZ case.} \label{fig7}
\end{figure}

It would be interesting to have a look at the temporal correlator in
the pure Gribov-Zwanziger case ($M^2=0$). Therefore,
$\mathcal{C}(t)$ is displayed in FIG.~\ref{fig7}. We can conclude
that this plot is grosso modo the same as in the refined
Gribov-Zwanziger setup.

\section{Conclusion}
In this paper, we have extended the $4D$ analysis of the
Gribov-Zwanziger scenario to $3D$. The motivations for our work were
to include an additional nonperturbative contribution to the
Gribov-Zwanziger action, related to a dynamical mass generation
effect, and the recent lattice data of
\cite{Cucchieri:2007md,Cucchieri:2007rg}. Two of the most striking
features of these simulations were a finite nonzero value of the
gluon propagator at zero momentum and the non-enhancement of the
ghost, in both $3D$ and $4D$ cases.\\

We have started from the following renormalizable action,
\begin{eqnarray}
S_{\GZ} +  M^2 \int \d^d x \left[ -\left( \overline{\varphi}^a_i
\varphi^a_{i} - \overline{\omega}^a_i \omega^a_i \right)  + 2
\frac{d (N^2 -1)}{\sqrt{2 g^2 N}}  \varsigma \ \gamma^2 \right] \;,
\end{eqnarray}
which is an extension of the ordinary Gribov-Zwanziger action
$S_{\GZ}$. Next to the local composite operator \lco\ we have also
added an extra vacuum term, necessary to ensure the restriction to
the Gribov region $\Omega$. In addition, the two parameters $\gamma$
and $\varsigma$ are determined by the following gap equations,
\begin{align}
\frac{\p \Gamma}{\p \gamma^2} & = 0\;,  &  \left.\frac{\partial
\sigma(0)}{\partial M^2} \right|_{M^2=0}&= 0\;.
\end{align}

Before fixing $M^2$, we have written down the explicit expression
for the ghost and the gluon propagator. Firstly, the one loop ghost
propagator $\mathcal{G}(k^2)$ is given by,
\begin{eqnarray}
\mathcal{G}(k^2) &=&  \frac{1}{k^2} \frac{1}{1- \sigma(k^2)}\;,
\end{eqnarray}
with $\sigma(k^2)$ given by expression \eqref{sigmakuitgerekend}. At
zero momentum, $\sigma(0)$ is given by
\begin{eqnarray}
    \sigma(0) &=& 1 -  \frac{M^4}{\lambda^4}\;,
\end{eqnarray}
already after applying the gap equations. We see that for $M^2
\not=0$, $\sigma(0)$ is smaller than 1, resulting in a non-enhanced
ghost propagator. Secondly, the tree level gluon propagator
$\mathcal{D}(p^2)$ yields,
\begin{equation}
    \mathcal{D}(p^2)=\frac{p^2 + M^2}{p^4 + M^2p^2 + \lambda^4}\;,
\end{equation}
with $\lambda^4 = 2 g^2 N \gamma^4$. Therefore, already at tree
level, we find that $\mathcal{D}(0) = \frac{M^2}{\lambda^4}$, which
is nonzero for $M^2 \not= 0$. We have also presented the one loop
gluon propagator at zero momentum in
equation \eqref{dnul}.\\

In the last part of this paper, we have discussed a variational
method in order to obtain an explicit value for $M^2$ in the
Gribov-Zwanziger model, which has provided a value to the zero
momentum ghost propagator (or equivalently $\sigma(0)$) and gluon
propagator. With this method, we have found
\begin{align}
\sigma(0) &= 0.94\;, & \mathcal{D}(0) &=
\frac{12}{\mathrm{GeV}^2}\;.
\end{align}
Both values are in qualitative agreement with the lattice data. With
this variational method, we also demonstrated the positivity
violation of the gluon propagator, which is also confirmed by
lattice data.\\

Let us also spend a few words on the applicability of a weak
coupling expansion. The reader will appreciate that we are no longer
perturbing around a trivial vacuum, but are in fact considering a
perturbative expansion around a nonperturbative vacuum characterized
by a nonvanishing Gribov mass $\lambda$, and additionally $M^2\neq0$
in our refined framework. These parameters ensure that the no-pole
condition \eqref{nopole} is fulfilled, which puts a nonperturbative
restriction on the theory. Obviously, we do not claim that we have
all the relevant nonperturbative dynamics enclosed in this
formalism, but at least the results we have do qualitatively match
the available lattice data. The validity of a systematic
perturbative expansion around this vacuum state is reflected in a
coupling
constant which should be sufficiently small, see e.g. \eqref{exppar}. This reasoning also applies to the $4D$ case, see \cite{Dudal:2008sp}.\\

In summary, we have presented a framework, which consistently
accounts for the recent large volume lattice data in the infrared
region in $3D$ and $4D$. The question is what happens in $2D$, as
the most recent data in $2D$ keeps predicting an enhanced ghost in
combination with a vanishing zero momentum gluon propagator,
contrasting with the higher-dimensional cases
\cite{Cucchieri:2007rg,Cucchieri:2008fc,Maas:2007uv}. Details of the
Gribov-Zwanziger framework in $2D$ shall be presented elsewhere, as
the situation is rather different there \cite{nieuw}.

Previous studies on the infrared behavior of the gluon and ghost
propagators within the Schwinger-Dyson formalism in $4D$ which where
in compliance with the lattice data can be found in
\cite{Boucaud:2006pc,Boucaud:2007va,Aguilar:2004sw}. In particular,
we observe that in \cite{Aguilar:2004sw} a gluon propagator fit was
given as $\mathcal D(p^2)=\frac{p^2+m_0^2}{p^4+m_0^2p^2+m_0^4}$,
while the ghost propagator almost behaved like $\frac{1}{p^2}$. It
is interesting to notice that this kind of gluon propagator is of
the same type as the one found here, in a completely different way.
Finally, let us also mention that similar research in the maximal
Abelian gauge \cite{Capri:2008ak} also gave results in qualitative
agreement with the available lattice data in that gauge, see
\cite{Mendes:2006kc}.\\

\section{Acknowledgments.}
We are grateful to A.~Cucchieri and T.~Mendes for providing us with
lattice data and useful comments. D.~Dudal and N.~Vandersickel are
supported by the Research Foundation - Flanders (FWO). The Conselho
Nacional de Desenvolvimento Cient\'{i}fico e Tecnol\'{o}gico
(CNPq-Brazil), the Faperj, Funda{\c{c}}{\~{a}}o de Amparo {\`{a}}
Pesquisa do Estado do Rio de Janeiro, the SR2-UERJ and the
Coordena{\c{c}}{\~{a}}o de Aperfei{\c{c}}oamento de Pessoal de
N{\'{i}}vel Superior (CAPES) are gratefully acknowledged for
financial support. This work is supported in part by funds provided
by the US Department of Energy (DOE) under cooperative research
agreement DEFG02-05ER41360.

\appendix

\section{Construction of the effective action $\Gamma(\sigma)$ from the generating functional $W(J)$\label{appendixA}}
\subsection{General framework}
We define the classical field $\sigma(x)$ conjugate to the source
$J(x)$ as follows\footnote{To avoid any confusion with
the reader, this classical field $\sigma(x)$ has obviously nothing to do with the one loop correction to the ghost form factor
$\sigma(k^2)$ first defined in equation \eqref{ghostpropagator1}.}:
\begin{equation}\label{inv1}
    \sigma(x)=\frac{\delta W(J)}{\delta J(x)} \;,
\end{equation}
with $W(J)$ the generating functional defined in \eqref{genfunc}.
The effective action $\Gamma(\sigma)$ is then obtained in the usual
way by a Legendre transformation
\begin{equation}\label{inv2}
    \Gamma(\sigma)= W(J)-\int \d^d x J(x)\sigma(x) \;,
\end{equation}
where $J(x)$ is understood to be a functional of $\sigma(x)$. It is
easily derived that
\begin{equation}\label{inv2bis}
    \frac{\delta\Gamma}{\delta\sigma(x)}= -J(x) \;.
\end{equation}
The original theory is recovered when the source $J$ attains again a
zero value. A source is nothing more than a tool to probe the
theory. In our case, we are incorporating the effects induced by the
operator $\overline{\varphi}\varphi-\overline{\omega}\omega$. At the
end, a source should always become zero. The equation
\eqref{inv2bis} for example corresponds to
the quantum version of the classical equation of motion when $J=0$.\\

 In what follows, we shall limit ourselves to
constant $J$ and $\sigma$, as we are mainly interested in the (space
time independent) vacuum expectation value of the operator coupled
to the source $J$. The remaining task is to obtain the functional
form of $\Gamma$ in terms of $\sigma$. Usually, this is done using
the background field formalism of Jackiw \cite{Jackiw:1974cv}, when
an effective potential $\Gamma(\phi_c)$ associated with an
elementary field $\phi(x)$ is determined, whereby it is understood
that $\phi(x)=\phi_c + \widetilde{\phi}(x)$. The quantity
$\widetilde{\phi}(x)$ represents the quantum fluctuations around the
vacuum expectation value of $\phi(x)$, i.e.
$\braket{\phi(x)}=\phi_c$, $\braket{\widetilde{\phi}(x)}=0$.
Unfortunately, when a composite operator is considered, this
procedure is less useful, as there is no elementary field associated
to the operator to begin with. Sometimes, techniques are at one's
disposal to introduce such a field, e.g. by employing a
Hubbard-Stratonovich transformation
\cite{Knecht:2001cc}.\\

 Let us proceed thus by explicitly performing the Legendre
transformation, along the lines of
\cite{Okumura:1994ec,Yokojima:1995hy}. The generating functional
$W(J)$ will be obtained as a series in the coupling constant,
\begin{equation}\label{inv3}
    W(J)=W_0(J)+g^2W_1(J)+\ldots \;.
\end{equation}
As a consequence, we may write
\begin{eqnarray}\label{inv5}
  \sigma(J) &=&\sigma_0(J)+g^2\sigma_1(J)+\ldots\;,\qquad  \mbox{with}\;\;\sigma_n(J)=\frac{\p W_n}{\p J} \;.
\end{eqnarray}
This last series can be inverted to give $J$ as a function of
$\sigma$,
\begin{eqnarray}
  J(\sigma) &=&J_0(\sigma)+g^2J_1(\sigma)+\ldots\;.
\end{eqnarray}
As a trivial consequence,
\begin{eqnarray}
 \sigma\equiv\sigma( J(\sigma)) &=&\sigma(J_0)+g^2\left(J_1\sigma_0^\prime(J_0)+\sigma_1(J_0)\right)+\ldots \;.
\end{eqnarray}
Since the field $\sigma$ is supposed to be independent of the
coupling constant at the level of the inversion\footnote{Only after
that the gap equation, $\frac{\p \Gamma}{\p \sigma}=0$, is imposed,
$\sigma$ will collect a $g^2$-dependence.}, we find an iterative
inversion procedure,
\begin{eqnarray}\label{invopl}
  \sigma = \sigma_0(J_0)&\Rightarrow& J_0=J_0(\sigma_0) \;,\nonumber\\
  0 =J_1\sigma_0^\prime(J_0)+\sigma_1(J_0) &\Rightarrow& J_1=-\frac{\sigma_1(J_0)}{\sigma_0^\prime(J_0)} \;, \nonumber\\
  &\vdots&
\end{eqnarray}
In this fashion, every term in the series for $J$ can be expressed
in terms of $J_0$, which itself is a function of $\sigma_0$.
Summarizing, everything can be written in terms of $\sigma_0$. Doing
so, we can calculate the r.h.s. of \eqref{inv2} up to the desired
order in $g^2$ by simply substituting the
corresponding expressions for $J_i$ in the l.h.s..\\

Once the inversion is performed, we must fix $\sigma_0$, which
corresponds to the condensate of the operator coupled to $J$, by
demanding that
\begin{equation}
    \frac{\p \left\{\Gamma_0(\sigma_0)+\Gamma_1(\sigma_0)+\ldots\right\}}{\p \sigma_0}=0 \Leftrightarrow    J_0(\sigma_0)+J_1(\sigma_0)+\ldots=0\;.
\end{equation}

\subsection{Application to the LCO $\overline{\varphi}\varphi-\overline{\omega}\omega$}
We shall now employ the previous method on the $3D$ Gribov-Zwanziger
action, in the presence of the LCO
\mbox{$\overline{\varphi}\varphi-\overline{\omega}\omega$}. The
action we use is given by \eqref{nact}. Using \eqref{inv5} and
\eqref{invopl}, yields
\begin{equation}\label{sigma}
    \sigma_0(J_0)=\frac{N^2-1}{6\pi}\left\{\frac{3}{2}\sqrt{J_0}-\frac{3}{4}\left(\frac{J_0+\sqrt{J_0^2-4\lambda^4}}{2}\right)\left(1+\frac{J_0}{\sqrt{J_0^2-4\lambda^4}}\right)
    -\frac{3}{4}\left(\frac{J_0-\sqrt{J_0^2-4\lambda^4}}{2}\right)\left(1-\frac{J_0}{\sqrt{J_0^2-4\lambda^4}}\right)\right\} \;.
\end{equation}
Defining $\widetilde{\sigma}\equiv\frac{6\pi}{N^2-1}\sigma_0$, we
found 6 different branches for the inverse function
$J_0(\widetilde{\sigma})$. However, only one of these was real
valued and gave rise to a solution of the gap equation, so we focus
our attention on this particular solution, \footnotesize
\begin{eqnarray*}
&& J_0(\widetilde{\sigma})
~=~\frac{36\lambda^2\wsigma^2-\wsigma^4}{27\wsigma^2}+\frac{\sqrt[3]{\wsigma^4\left[8192\wsigma^{8}+442368\lambda^2\wsigma^{6}+6469632\lambda^4\wsigma^4+20901888\lambda^6\wsigma^2
 +45349632\lambda^8+186624\sqrt{3}\lambda^5(9\lambda^2+4\wsigma^2)\sqrt{243\lambda^2+8\wsigma^2}\right]}}{432\sqrt[3]{2}\wsigma^2} \nonumber\\
 &&+ \frac{256\wsigma^8+9216\lambda^2\wsigma^6+51840\lambda^4\wsigma^4}{2162^{2/3}\wsigma^2 \sqrt[3]{\wsigma^4\left[8192\wsigma^{8}+442368\lambda^2\wsigma^{6}+6469632\lambda^4\wsigma^4+20901888\lambda^6\wsigma^2
 +45349632\lambda^8+186624\sqrt{3}\lambda^5(9\lambda^2+4\wsigma^2)\sqrt{243\lambda^2+8\wsigma^2}\right]}} \;.
\end{eqnarray*}
\normalsize
\begin{figure}[t]
\begin{center}
  \scalebox{1}{\includegraphics{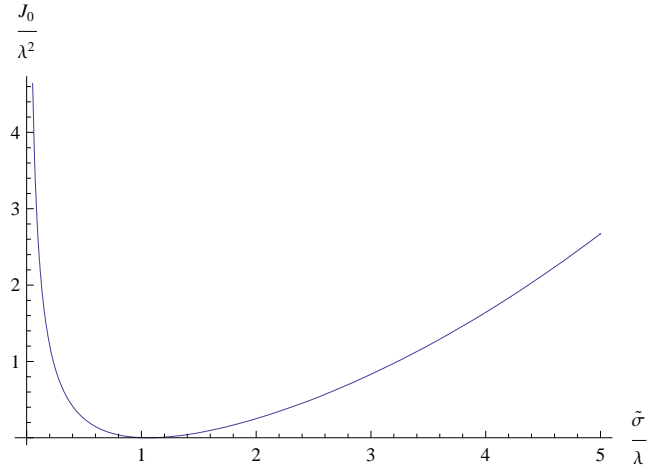}}
\caption{A plot of $\frac{J_0}{\lambda^2}$ in terms of
$\frac{\wsigma}{\lambda}$.}\label{fig9}
\end{center}
\end{figure}
We have displayed a plot of $J_0/\lambda^2$ in terms of
$\wsigma/\lambda$ in FIG.\ref{fig9}. Solving the gap equation
$J_0(\wsigma)=0$ numerically leads to
\begin{equation}\label{oplgap}
    \sigma_0\approx 1.06\frac{N^2-1}{6\pi}\lambda\approx0.056(N^2-1)\lambda \;.
\end{equation}
This numerical solution is consistent with the analytically derived
solution \eqref{pot5}. Let us now determine the Gribov mass
$\lambda$. Therefore, we first compute the effective action using
its definition \eqref{inv2}. We did not write down the eventual
result in terms of $\lambda$ and $\wsigma$, as the expression is
quite lengthy. We subsequently solved the gap equation (horizon
condition) $\frac{\p \Gamma}{\p\lambda}=0$ (see FIG.\ref{fig10})
numerically in the case $N=3$, and we found
\begin{equation}\label{gribovopl}
    \lambda\approx 0.113 g^2 \;,
\end{equation}
which is also consistent with the analytical result
\eqref{gribovpuuropl}. The corresponding vacuum energy is given by
\begin{equation}\label{gribovvac}
    E_{\mathrm{vac}}\approx 0.000214g^6 \;,
\end{equation}
again equivalent with \eqref{gribovpuurvac}.\\

The gap equation derived from the effective action
$\Gamma(\sigma_0,\lambda)$ should of course give back this
perturbative solution, next to a potential nonperturbative solution,
which can never be discovered by simply using \eqref{pot4}. The
whole point of the inversion (Legendre transformation) is just that
there might be multiple solutions to the equation $J=0$, next to the
perturbative one. In our case, the situation is interesting because
there is already a nontrivial scale in the game, namely the Gribov
mass $\lambda$. This
 allows us to obtain a nonvanishing value for the
condensate already at the perturbative level.
\begin{figure}[t]
\begin{center}
  \scalebox{1}{\includegraphics{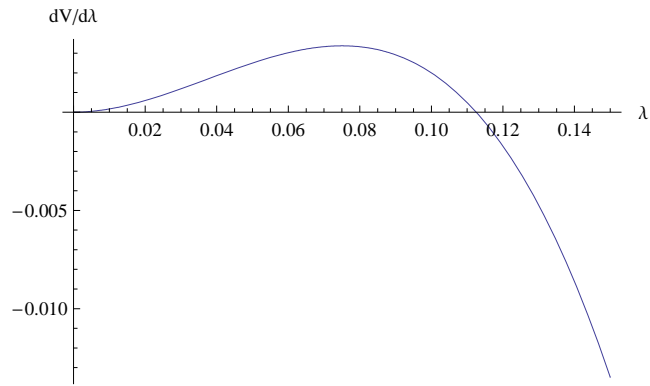}}
\caption{The horizon function $\frac{\p \Gamma}{\p \lambda}$ for
$N=3$ in units $g^2=1$.}\label{fig10}
\end{center}
\end{figure}

\section{The renormalizability of the extended Gribov-Zwanziger action: algebraic analysis}
\subsection{The Ward identities}
The action \eqref{cact} exhibits several Ward identities which we
have enlisted below.
\begin{itemize}
\item The global $U(f)$ invariance:
\begin{eqnarray}
U_{ij} \Sigma &=&0\;, \label{ward1} \\
U_{ij}&=&\int \d^{d}x\left( \varphi _{i}^{a}\frac{\delta
}{\delta\varphi_{j}^{a}}-\overline{\varphi}_{j}^{a}\frac{\delta
}{\delta\overline{\varphi}_{i}^{a}}+\omega _{i}^{a}\frac{\delta
}{\delta \omega _{j}^{a}}-\overline{\omega }_{j}^{a}\frac{\delta
}{\delta\overline{\omega}_{i}^{a}}\right) \nonumber\;.
\end{eqnarray}

\item  The Slavnov-Taylor identity:
\begin{equation}
\mathcal{S}(\Sigma )=0\;,
\end{equation}
\begin{eqnarray}
\mathcal{S}(\Sigma ) &=&\int \d^{d}x\left( \frac{\delta \Sigma
}{\delta K_{\mu }^{a}}\frac{\delta \Sigma }{\delta A_{\mu
}^{a}}+\frac{\delta \Sigma }{\delta L^{a}}\frac{\delta \Sigma
}{\delta c^{a}}+b^{a}\frac{\delta \Sigma
}{\delta \overline{c}^{a}}+\overline{\varphi }_{i}^{a}\frac{\delta \Sigma }{%
\delta \overline{\omega }_{i}^{a}}+\omega _{i}^{a}\frac{\delta
\Sigma }{\delta \varphi _{i}^{a}}+M_{\mu }^{ai}\frac{\delta \Sigma
}{\delta U_{\mu}^{ai}}+N_{\mu }^{ai}\frac{\delta \Sigma }{\delta
V_{\mu }^{ai}}\right) \;. \nonumber
\end{eqnarray}

\item  The Landau gauge condition and the antighost equation:
\begin{eqnarray}
\frac{\delta \Sigma }{\delta b^{a}}&=&\partial_\mu A_\mu^{a}\;,
\label{r11}\\ \frac{\delta \Sigma }{\delta
\overline{c}^{a}}+\partial _{\mu }\frac{\delta \Sigma }{\delta
K_{\mu }^{a}}&=&0\;.
\end{eqnarray}

\item  The ghost Ward identity:
\begin{eqnarray}
\mathcal{G}^{a}\Sigma &=&\Delta _{\mathrm{cl}}^{a}\;, \\
\mathcal{G}^{a} &=&\int \d^{d}x\left( \frac{\delta }{\delta c^{a}}+gf^{abc}\left( \overline{c}^{b}\frac{\delta }{\delta b^{c}}+\varphi _{i}^{b}\frac{\delta }{\delta \omega _{i}^{c}} +\overline{\omega }_{i}^{b}\frac{\delta }{\delta \overline{\varphi }_{i}^{c}} +V_{\mu }^{bi}\frac{\delta }{\delta N_{\mu }^{ci}}+U_{\mu }^{bi}\frac{\delta }{\delta M_{\mu }^{ci}}\right) \right) \;,  \nonumber \\
\Delta _{\mathrm{cl}}^{a}&=& g\int \d^{d}xf^{abc}\left(
K_{\mu}^{b}A_{\mu }^{c}-L^{b}c^{c}\right) \;.  \nonumber
\end{eqnarray}
Notice that $\Delta _{\mathrm{cl}}^{a}$ is a classical breaking,
since it is linear in the quantum fields.

\item  The linearly broken local constraints:
\begin{eqnarray}
&&\frac{\delta \Sigma }{\delta \overline{\varphi }^{ai}}+\partial _{\mu }\frac{\delta \Sigma }{\delta M_{\mu }^{ai}}=gf^{abc}A_{\mu }^{b}V_{\mu}^{ci}+J\varphi_i^a\;, \\
&&\frac{\delta \Sigma }{\delta \omega ^{ai}}+\partial
_{\mu}\frac{\delta \Sigma }{\delta N_{\mu
}^{ai}}-gf^{abc}\overline{\omega }^{bi}\frac{\delta \Sigma }{\delta b^{c}}=gf^{abc}A_{\mu }^{b}U_{\mu }^{ci}+J\overline{\omega}_i^a\;, \\
&&\frac{\delta \Sigma }{\delta \overline{\omega }^{ai}}+\partial _{\mu }\frac{%
\delta \Sigma }{\delta U_{\mu }^{ai}}-gf^{abc}V_{\mu
}^{bi}\frac{\delta \Sigma }{\delta K_{\mu }^{c}}=-gf^{abc}A_{\mu
}^{b}N_{\mu }^{ci} -J\omega_i^a\;,  \\
&&\frac{\delta \Sigma }{\delta \varphi ^{ai}}+\partial
_{\mu}\frac{\delta \Sigma }{\delta V_{\mu
}^{ai}}-gf^{abc}\overline{\varphi }^{bi}\frac{\delta \Sigma }{\delta
b^{c}}-gf^{abc}\overline{\omega }^{bi}\frac{\delta \Sigma }{\delta
\overline{c}^{c}} -gf^{abc}U_{\mu }^{bi}\frac{\delta \Sigma}{\delta
K_{\mu }^{c}}  =gf^{abc}A_{\mu }^{b}M_{\mu
}^{ci}+J\overline{\varphi}_i^a\;.
\end{eqnarray}

\item  The exact $\mathcal{R}_{ij}$ symmetry:
\begin{eqnarray}
\mathcal{R}_{ij}\Sigma &=&0\;,  \label{ward7}\\
\mathcal{R}_{ij}&=&\int \d^{d}x\left( \varphi
_{i}^{a}\frac{\delta}{\delta\omega _{j}^{a}}-\overline{\omega
}_{j}^{a}\frac{\delta }{\delta \overline{\varphi }_{i}^{a}}-V_{\mu
}^{ai}\frac{\delta }{\delta N_{\mu }^{aj}}+U_{\mu }^{aj}\frac{\delta
}{\delta M_{\mu }^{ai}}\right) \;.\nonumber
\end{eqnarray}
\end{itemize}
We shall now rely on the algebraic renormalization procedure
\cite{Piguet:1995er}, according to which the most general allowed
 invariant counterterm $\Sigma ^{c}$ is an integrated
local polynomial in the fields and sources with dimension bounded by
three, with vanishing ghost number and $Q_{f}$-charge, and subject
to the following constraints:
\begin{eqnarray}\label{conditiebegin}
U_{ij} \Sigma^c &=&0 \;, \nonumber\\
\mathcal{B}_{\Sigma }\Sigma ^{c}&=&0\;,\nonumber\\
\frac{\delta \Sigma }{\delta \overline{c}^{a}}+\partial _{\mu}\frac{\delta\Sigma }{\delta K_{\mu }^{a}} &=&0\;,  \nonumber \\
\frac{\delta \Sigma ^{c}}{\delta b^{a}} &=&0\;,\nonumber\\
\mathcal{G}^{a}\Sigma ^{c}&=&0\,,\nonumber\\
\frac{\delta \Sigma ^{c}}{\delta \varphi ^{ai}}+\partial _{\mu}\frac{\delta\Sigma ^{c}}{\delta V_{\mu }^{ai}}-gf^{abc}\overline{\omega }^{bi}\frac{\delta \Sigma ^{c}}{\delta \overline{c}^{c}}-gf^{abc}U_{\mu }^{bi}\frac{\delta \Sigma ^{c}}{\delta K_{\mu }^{c}} &=&0\;,  \nonumber \\
\frac{\delta \Sigma ^{c}}{\delta \overline{\omega }^{ai}}+\partial _{\mu }\frac{\delta \Sigma ^{c}}{\delta U_{\mu }^{ai}}-gf^{abc}V_{\mu }^{bi}\frac{\delta \Sigma ^{c}}{\delta K_{\mu }^{c}} &=&0\;,  \nonumber \\
\frac{\delta \Sigma ^{c}}{\delta \omega ^{ai}}+\partial _{\mu}\frac{\delta \Sigma ^{c}}{\delta N_{\mu }^{ai}} &=&0\;,\nonumber\\
\frac{\delta \Sigma }{\delta \overline{\varphi }^{ai}}+\partial
_{\mu }\frac{\delta \Sigma }{\delta M_{\mu }^{ai}} &=&0\;,
\nonumber\\\mathcal{R}_{ij}\Sigma ^{c}&=&0\;.
\end{eqnarray}
The operator $B_\Sigma$ is the nilpotent linearized Slavnov-Taylor
operator,
\begin{eqnarray}
 \mathcal{B}_{\Sigma } &=&\int \d^{4}x\left(
\frac{\delta \Sigma}{\delta K_{\mu }^{a}}\frac{\delta }{\delta
A_{\mu }^{a}}+\frac{\delta \Sigma }{\delta A_{\mu }^{a}}\frac{\delta
}{\delta K_{\mu }^{a}}+\frac{\delta \Sigma }{\delta
L^{a}}\frac{\delta }{\delta c^{a}}+\frac{\delta\Sigma }{\delta
c^{a}}\frac{\delta }{\delta L^{a}}+b^{a}\frac{\delta }{\delta
\overline{c}^{a}}+ \overline{\varphi}_{i}^{a}\frac{\delta }{\delta
\overline{\omega }_{i}^{a}}+\omega_{i}^{a}\frac{\delta }{\delta
\varphi_{i}^{a}}+M_{\mu }^{ai}\frac{\delta }{\delta U_{\mu
}^{ai}}+N_{\mu }^{ai}\frac{\delta }{\delta V_{\mu }^{ai}} \right)
\;,\nonumber\\ \mathcal{B}_{\Sigma }\mathcal{B}_{\Sigma }&=&0\;.
\end{eqnarray}
One can show that $\Sigma ^{c}$ does not depend on the Lagrange
multiplier $b^{a}$, and that the antighost $\overline{c}^{a}$ and
the $i$-valued fields $\varphi_{i}^{a}$, $\omega _{i}^{a}$,
$\overline{\varphi }_{i}^{a}$, $\overline{\omega }_{i}^{a}$ can only
enter through the combinations \cite{Zwanziger:1992qr}
\begin{align}
&\widetilde{K}_{\mu }^{a} = K_{\mu }^{a}+\partial _{\mu }\overline{c}^{a}-gf^{abc}\widetilde{U}_{\mu }^{bi}\varphi ^{ci}-gf^{abc}V_{\mu }^{bi}\overline{\omega }^{ci}\;,  \nonumber \\
&\widetilde{U}_{\mu }^{ai} =U_{\mu }^{ai}+\partial _{\mu
}\overline{\omega }^{ai}\;,
\widetilde{V}_{\mu }^{ai}  =V_{\mu }^{ai}+\partial _{\mu }\varphi^{ai}\;,\nonumber \\
&\widetilde{N}_{\mu }^{ai} =N_{\mu }^{ai}+\partial _{\mu
}\omega^{ai}\;, \widetilde{M}_{\mu }^{ai}  =V_{\mu }^{ai}+\partial
_{\mu}\overline{\varphi }^{ai}\;.
\end{align}
Imposing the previous constraints arising from the Ward identities,
the most general counterterm can be brought in the following compact
form
\begin{eqnarray}
\Sigma ^{c}&=&a_{0}S_{YM}+a_{1}\int
\d^{d}x\left(A_{\mu}^{a}\frac{\delta S_{YM}}{\delta
A_{\mu}^{a}}+\widetilde{K}_{\mu }^{a}\partial _{\mu }c^{a}+
\widetilde{V}_{\mu}^{ai}\widetilde{M}_{\mu }^{ai}
-\widetilde{U}_{\mu }^{ai}\widetilde{N}_{\mu }^{ai}\right)+a_2\int
\d^d x \rho g^2J \;, \label{counterterm}
\end{eqnarray}
with $a_{0}$, $a_{1}$ and $a_2$ three  arbitrary parameters. It is
an impressive feature of the action $\Sigma$ that only 3 independent
parameters can enter the counterterm, despite the
 presence of many fields and sources.

\subsection{Stability of the action and the renormalization constants.}
As the most general counterterm $\Sigma ^{c}$ compatible with the
Ward identities has now been constructed, it still remains to be
checked whether the starting action $\Sigma $ is stable, i.e. that
$\Sigma ^{c}$ can be reabsorbed into $\Sigma$ through a
renormalization of the parameters, fields and sources appearing in
$\Sigma $.\\

According to expression (\ref{counterterm}), $\Sigma ^{c}$ contains
three parameters $a_{0}$,$a_{1}$ and $a_2$, which correspond to a
multiplicative renormalization of the gauge coupling constant $g$,
the parameter $\rho $, the fields $\phi =(A_{\mu }^{a}$, $c^{a}$,
$\overline{c}^{a}$, $b^{a}$, $\varphi _{i}^{a}$, $ \omega _{i}^{a}$,
$\overline{\varphi }_{i}^{a}$, $\overline{\omega }_{i}^{a}) $ and
the sources $\Phi =(K^{a\mu }$, $L^{a}$, $M_{\mu }^{ai}$, $ N_{\mu
}^{ai}$, $V_{\mu }^{ai}$, $U_{\mu }^{ai},J)$, according to
\begin{equation}
\Sigma (g,\rho ,\phi ,\Phi )+\eta \Sigma
^{c}=\Sigma(g_{0},\rho_{0},\phi _{0},\Phi _{0})+O(\eta ^{2})\;.
\label{co1}
\end{equation}
This is possible, provided that we define the bare quantities in
terms of the renormalized quantities as
\begin{align}
g_{0}&=Z_{g}g\;, & \rho _{0}&=Z_{\rho }\rho \;, \nonumber\\
\phi _{0} &=Z_{\phi}^{1/2}\phi \;, & \Phi _{0} &=Z_{\Phi }\Phi \;,
\end{align}
with
\begin{align}
Z_{g} &=1+\eta \frac{a_{0}}{2} \;, & Z_{A}^{1/2} &=1+\eta
\left(a_{1}-\frac{a_{0}}{2}\right)\;,&Z_\rho&=1+\eta\left(a_2-a_1-a_0\right)
\;.
\end{align}
These are the only independent renormalization constants. For the
rest of the fields, we have
\begin{equation}\label{renrel2}
Z_{c}=Z_{\overline{c}}=Z_{\varphi }=Z_{\overline{\varphi
}}=Z_{\omega }=Z_{\overline{\omega }} =Z_{g}^{-1}Z_{A}^{-1/2}\;,
\end{equation}
while for the renormalization of the sources
$\left(M_{\mu}^{ai},N_{\mu }^{ai},V_{\mu }^{ai},U_{\mu
}^{ai},J\right)$
\begin{align}
Z_{M}&=Z_{N}=Z_{V}=Z_{U}=Z_{g}^{-1/2}Z_{A}^{-1/4}\;,& Z_J&=Z_g
Z_A^{1/2} \;. \label{renrel}
\end{align}
We see thus that the LCO
$\overline{\varphi}\varphi-\overline{\omega}\omega$ does not
renormalize independently, as it is evident from \eqref{renrel}. The
only new parameter entering the game corresponds to the
renormalization of the vacuum functional, expressed by $\rho$ and
its renormalization factor $Z_\rho$. As discussed in the main body
of this paper, this parameter turns out to be equal to zero anyhow.

\section{Propagators in lattice and continuum formulation}
 For the benefit of the reader, in this Appendix we
discuss how our results compare with the corresponding lattice
data in the case of the gluon propagator. The quantity which is
evaluated in both cases is the gluon propagator, namely the
connected gluon two-point function $\Braket{A_\mu^a(x)
A_\nu^b(y)}$, where the gauge field configurations $A_\mu^a$ are
restricted to the Gribov region $\Omega$.  We shall also show that the gluon and the ghost propagators are color diagonal.

\subsection{Continuum formulation}
In the continuum formulation, the gluon propagator is given
by the connected gluon two-point function, and expressible by means
of
\begin{eqnarray}\label{prop1}
\Braket{A_\mu^a(x) A_\nu^b(y) } &=& \left.\frac{\delta^2
Z^c(J)}{\delta J^a_\mu (x) \delta J^b_\nu (y) }\right\vert_{J=0}\,,
\end{eqnarray}
with $Z^c(J)$ the generating functional of the connected gluon
$n$-point functions, which in our case will read
\begin{eqnarray}\label{prop3}
\e^{-Z^c(J)}~\equiv~ \e^{-Z^c(J,J_\varphi,J_{\overline{\varphi}})}
~=~ \int [\d \Psi] \e^{-\left(S_{\tot} + \int \d^4 x \left(J^a_\mu
A^a_\mu+J_{\varphi, \mu}^{ab} \varphi_\mu^{ab}
+J_{\overline{\varphi}, \mu}^{ab}\overline{\varphi}_\mu^{ab}\right)\right)}\,,
\end{eqnarray}
where $S_{\tot}$ is the improved Gribov-Zwanziger action
\eqref{totaleactie}. This amounts to considering the Landau gauge
fixing, such that the relevant gluon configurations belong to the
Gribov region, i.e. these are
(local) minima of $\int \d^3x A^2$ along the gauge orbit.\\
\\
As proven in \cite{Dudal:2008sp}, the gluon propagator \eqref{prop1}
is transverse.  In a
condensed notation, one shall find
\begin{eqnarray}\label{prop4}
\e^{-Z^c(J,J_\varphi,J_{\overline{\varphi}})} &=& \e^{-\left(
                                            \begin{array}{ccc}
                                              J & J_\varphi & J_{\overline{\varphi}} \\
                                            \end{array}
                                          \right)
 {\cal M}
\left(\begin{array}{c}
                                                            J \\
                                                            J_\varphi \\
                                                            J_{\overline{\varphi}}
                                                            \end{array}\right)\,+\,\mbox{higher order terms in $J$, $J_\varphi$,
                                                            $J_{\overline{\varphi}}$}}\,,
\end{eqnarray}
where ${\cal M}$ is the matrix propagator, as written down in
\eqref{matrix} up to first order. The upper left corner of this
matrix ${\cal M}$ corresponds precisely to the gluon propagator,
 as it is apparent by taking the second derivative with
respect to
the source $J^a_\mu$. \\\\\
If one is interested in the 1PI two-point functions, one should look
at the corresponding generator, which is the effective action
$\Gamma[A_\mu,\varphi,\overline{\varphi}]$. As is well known, the
corresponding 1PI two-point function will be the inverse of the
connected two-point function (or propagator). Said otherwise, the
corresponding matrices will be each others inverse. This is also
explained in the main body of the text, with the 1PI two-point
function matrix written
down in \eqref{1PImatrix}, again up to first order.\\\\
As we have already stressed earlier in the paper, the extra fields
$\left(\overline{\varphi}_{\mu}^{ac},\varphi_{\mu}^{ac},\overline{\omega}_{\mu
}^{ac},\omega_{\mu}^{ac}\right)$ are introduced in order to obtain a
local manageable field theory, which is capable of restricting the
gauge field configurations to the Gribov region, which is a rather
nontrivial operation in the continuum. In principle, one could opt
to work in an effective field theory fashion by again integrating out the extra fields. Clearly, this will give rise to a very
complicated (nonlocal) action, written solely in terms of the
original Yang-Mills fields. In this formulation, the gluon
propagator is directly related to the inverse of the 1PI two-point
function, due to the absence of mixing. Anyhow, the result for the
propagator itself will be the same as the one already obtained
before in the preferable local and manifestly renormalizable
formulation with the extra fields, when looking at the same order in
$g^2$. This can be easily checked at tree level: integrating out the
auxiliary fields in \eqref{totaleactie} leads to the following
quadratic (nonlocal) effective action,
\begin{eqnarray}\label{approx}  S_{\mathrm{quad}}&=&\int \d^3x \; \left[ \frac{1}{4} \left( \p_{\mu} A_{\nu}^a -
\p_{\nu}A_{\mu}^a \right)^2+ \frac{1}{2\alpha} \left( \p_{\mu}
A^a_{\mu} \right)^2  - N \gamma^4 g^2 A_{\mu}^a \frac{1}{\p^2 -M^2}
A^a_{\mu} + \ldots \right]\,,
\end{eqnarray}
 where the limit $\alpha
\rightarrow 0$ is understood in order to  recover the Landau gauge,
and where we skipped the irrelevant constant terms. The tree level
gluon propagator in momentum space is in this case $1/{\cal Q}_2$,
with ${\cal Q}_2$ the quadratic form appearing in \eqref{approx},
when expressed in momentum space. Clearly, this leads back to the
lowest order approximation in the upper left corner of
\eqref{matrix}.\\
\\
Concerning the ghost propagator, a similar formalism applies. \\
\\
Let us discuss the issue of color diagonality. The global color symmetry guarantees us that the gluon and the ghost propagators are color diagonal. This property is encoded in the global $SU(N)$ Ward identity, which reads at the classical level
\begin{equation}
    \int \d^d x \left( (\delta^{b}_{\mathrm{adj}}A_\mu^a)\frac{\delta\Sigma}{\delta A_\mu^a}+ \sum_\phi (\delta^b_{\mathrm{adj}} \phi) \frac{\delta \Sigma}{\delta \phi} \right) = 0 \;,
\end{equation}
with $\Sigma$ the classical action and $\phi$ all the other fields.
In particular,
\begin{eqnarray}
\delta^{b}_{\mathrm{adj}}A_\mu^a &=&  f^{abc} A^c_\mu\;,
\end{eqnarray}
and similar relations for the other fields $\phi$. Therefore, we find
\begin{equation}
    \int \d^d x\left(f^{abc}A_\mu^c\frac{\delta\Sigma}{\delta A_\mu^a}+\ldots\right) = 0 \;.
\end{equation}
This identity can be upgraded to the quantum level\footnote{We refer to \cite{Piguet:1995er} for the explicit proof.},
\begin{equation}
    \int \d^d x\left(f^{abc}A_\mu^c\frac{\delta\Gamma}{\delta A_\mu^a}+\ldots\right) = 0 \;,
\end{equation}
with $\Gamma$ the generator of 1PI correlators. Performing the Legendre transformation leads to the analogous Ward identity for the generator $Z^c$ of connected correlators,
\begin{equation}
    \int \d^d x\left(f^{abc}J_\mu^a\frac{\delta Z^c}{\delta J_\mu^c}\right) = 0\;.
\end{equation}
We shall concentrate on the gluon sector, so we have already set all other sources equal to zero. Taking a derivative w.r.t.~$J_\kappa^d(y)$ leads to
\begin{equation}
    f^{dbc}\frac{\delta Z^c}{\delta J_\kappa^c(y)}     + \int \d^d x\left(f^{abc}J_\mu^a(x)\frac{\delta^2 Z^c}{\delta J_\mu^c(x)\delta J_\kappa^d(y)}\right) = 0\;.
\end{equation}
Next, taking a derivative w.r.t. $J_\lambda^\ell(z)$ and setting $J=0$ at the end gives the following relationship,
\begin{equation}
    f^{dbc}\left.\frac{\delta^2 Z^c}{\delta J_\lambda^\ell(z)\delta
    J_\kappa^c(y)}\right\vert_{J=0}  +       f^{\ell bc}\left.\frac{\delta^2 Z^c}{\delta J_\lambda^c(z)\delta
    J_\kappa^d(y)}\right\vert_{J=0}  =0\;,
\end{equation}
or equivalently
\begin{eqnarray}
f^{dbc} \Braket{A^c_\kappa (y) A^\ell_\lambda(z)} + f^{\ell bc} \Braket{A^d_\kappa(y) A^c_\lambda(z)} &=&0\;.
\end{eqnarray}
This relation expresses nothing else than that the gluon propagator is an $SU(N)$ invariant rank two tensor.  Therefore,
\begin{eqnarray}
 \Braket{A^c_\kappa (y) A^\ell_\lambda(z)}  & \propto & \delta^{c \ell}\;,
\end{eqnarray}
since $\delta^{c \ell}$ is the unique invariant rank two tensor.\\
\\
Obviously, all available explicit loop results obtained with the (modified) Gribov-Zwanziger action are compatible with the general proof. Notice also that the proof is the same as the one we would
use in the case of normal $SU(N)$ gauge theories.\\
\\
An analogous result can be derived for the ghost propagator.

\subsection{Lattice formulation}
In a lattice formulation, one also calculates the
connected two-point function, by taking the Monte Carlo average of
the discrete version of the operator $\Braket{A_\mu^a(x) A_\nu^b(y)
}$. The statistical weight for this simulation is given by the
exponential of the discretized version of the Yang-Mills gauge
action, e.g. the Wilson action. The Landau gauge fixing is
numerically implemented by minimizing a suitable functional along
the gauge orbits, which corresponds to minimizing $\int \d^3 x
A^2$ in the continuum. As we have already explained in the
introduction, this amounts to numerically selecting a gauge
configuration within the Gribov region, equivalent
with what we did in the continuum. We refer the interested reader
to Section 2 of \cite{Bloch:2003sk} for the explicit expressions
of the discrete action, gauge fields and minimizing functional. In
particular, we refer to Subsection 2.4 in which the continuum and
lattice versions of the gluon propagator are written down.  The lattice gluon propagator also turns out to be transverse. Moreover, both the gluon and ghost propagator are found to be color diagonal. \\
\\
We emphasize here that lattice simulations thus never directly
calculate any 1PI two-point function, but, we repeat, do also
calculate the (connected) two-point correlator, i.e. the propagator
itself.

\end{document}